# The dynamics of viruslike capsid assembly and disassembly

Suzanne B. P. E. Timmermans,[1,∥] Alireza Ramezani,[2,∥] Toni Montalvo,[2] Mark Nguyen,[2] Paul van der Schoot,[3] Jan C. M. van Hest,[1*] and Roya Zandi[2*]

1 Bio-Organic Chemistry Research Group, Institute for Complex Molecular Systems, Eindhoven University of Technology, P.O. Box 513, 5600 MB Eindhoven, The Netherlands; 2 Department of Physics and Astronomy, University of California Riverside, California 92521, USA; 3 Soft Matter and Biological Physics Group, Department of Applied Physics, Eindhoven University of Technology, P.O. Box 513, 5600 MB Eindhoven, The Netherlands.

**ABSTRACT:**

Cowpea chlorotic mottle virus (CCMV) is a widely used model for virus replication studies. A major challenge lies in distinguishing between the roles of the interaction between coat proteins and that between the coat proteins and the viral RNA in assembly and disassembly processes. Here, we report on the spontaneous and reversible size conversion of the empty capsids of a CCMV capsid protein functionalized with a hydrophobic elastin-like polypeptide which occurs following a pH jump. We monitor the concentration of $T = 3$ and $T = 1$ capsids as a function of time and show that the time evolution of the conversion from one $T$ number to another is not symmetric: the conversion from $T = 1$ to $T = 3$ is a factor of 10 slower than that of $T = 3$ to $T = 1$. We explain our experimental findings using a simple model based on classical nucleation theory applied to virus capsids, in which we account for the change in the free protein concentration, as the different types of shells assemble and disassemble by shedding or absorbing single protein subunits. As far as we are aware, this is the first study confirming that both the assembly and disassembly of viruslike shells can be explained through classical nucleation theory, reproducing quantitatively results from time-resolved experiments.

Single-stranded RNA (ssRNA) viruses infect all species in the tree of evolution, causing significant economic damage and health concerns. The ssRNA genome of such viruses is protected by a shell called the capsid, composed of many copies of a single or a few protein subunits. To infect a host cell, a virus needs to enter, disassemble, release its genome, and use the cell's machinery for replication. Clearly, the capsid is a responsive structure: although it protects the genome and should be stable outside the cell, it must also readily disassemble once inside the cell and present its genome for replication.[1,2]

Arguably the most extensively studied viruses in this context are Cowpea Chlorotic Mottle Virus (CCMV) and Brome Mosaic Virus (BMV), which have proven to be good models for virus replication studies. The disassembly of the capsid in a cell must be triggered by changes in the chemical environment, resulting in the weakening of molecular interactions. Indeed, in vitro studies of CCMV and BMV show that following a pH jump from a neutral to a basic environment at high ionic strength, the capsids of these viruses spontaneously disassemble.[3–5] However, since the spatial and temporal resolution of intermediate structures of these studies are limited, kinetic pathways of disassembly have remained a mystery.

Generally, despite a huge body of work dedicated to understanding virus uncoating, our understanding of its kinetics and the factors contributing to it remains rudimentary.[6–15] One of the main reasons for the lack of insight is the fact that the *assembly* of CCMV is governed by two driving forces involving two species, namely, the interaction between the capsid proteins (CPs) and that between the ssRNA and the RNA-binding domain of CPs.[16] Distinguishing the contribution of both in the *disassembly* is not trivial, as CCMV shells in the absence of genome are not stable under physiological conditions.[17,18]

To develop and validate a plausible model that describes capsid assembly *and* disassembly, experimental conditions have to be found that allow for the elimination of the contribution of nucleic acids. This would not only lead to a better understanding of virus assembly but also allow for the development of tools to manipulate this process, either by preventing capsid formation and counteracting viral replication or by stabilizing empty capsids under physiological conditions as tools for diagnostic and therapeutic applications.[19]

Several years ago we designed the CP variant ELP-CP, which involves the attachment of elastin-like polypeptides (ELPs) at the N-terminus of the CPs of CCMV.[20] These ELPs consist of nine repeating Val-Pro-Gly-Xaa-Gly pentapeptide units, which switch from an extended water-soluble state to a collapsed hydrophobic state in response to an increase in temperature and/or electrolyte concentration.[21] The sequence contains 2 times the Trp, 2 times the Val, 4 times the Leu, and 1 time Gly as the guest residues (Xaa). At pH 5, the ELP-CPs form viruslike particles (VLPs) with a diameter of 28 nm, similar to the native $T = 3$ particles.[20] At pH 7.5, wild-type CPs do not assemble into shells, yet ELP-CPs assemble into 18 nm ($T=1$) VLPs upon

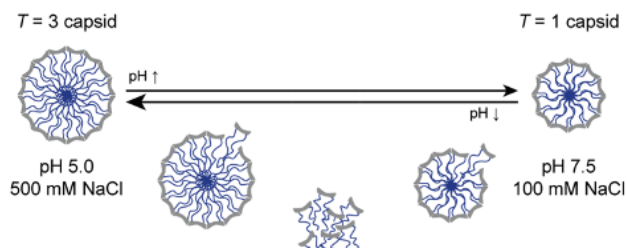

**Figure 1**. Schematic overview of the size change of ELP-CP viruslike particles (VLPs) upon a shift in pH.

increasing the salt concentration, a process induced by the hydrophobicity of the ELPs.[20,22,23]

In this paper, we describe the results from time-resolved experiments, allowing us to investigate the disassembly of one type of ELP-CP capsid and reassembly of another in response to pH changes (Figure 1). While changing the pH from 5 to 7.5, we monitor as a function of time how the $T = 3$ shells disappear, while the $T = 1$ shells appear. We also study the disassembly of $T = 1$ capsids and the assembly of $T = 3$ capsids by lowering the pH from 7.5 to 5. Our experimental findings can be explained by a simple model based on classical nucleation theory (CNT) applied to viruslike capsids,[24–27] accounting for the time-evolution of the concentrations of the various species that result from the shedding or addition of single protein subunits as the different types of shell assemble and disassemble. As far as we are aware, this is the first study confirming that both assembly and disassembly of viruslike shells can be explained through CNT as a possible mechanism for quantitatively reproducing experimental data.

For this purpose, we investigate the T number conversion over time, using a combination of size exclusion chromatography (SEC) and transmission electron microscopy (TEM). We first evaluate the conversion dynamics from $T = 3$ to $T = 1$ particles. Hereto, we dialyzed a 100 μM solution of empty VW1-VW8 ELP-CCMV $T = 3$ capsids at 4 °C from a pH 5.0 buffer with 500 mM NaCl to a pH 7.5 buffer with 100 mM NaCl, thus simultaneously increasing the pH and decreasing the ionic strength of the buffer environment. In order to stabilize the samples during SEC measurements, 0.2 equiv of $Ni^{2+}$ was added (Supporting Information section 2.3 for optimization). Details of our experimental procedures are found in Supporting Information Sections 1 and 2. As our experiments reveal that this process is very much dependent on the NaCl concentration in the buffer (Figures S1 and S2), we conclude that it must be driven by the stimulus-responsive ELP-domains. Because of this NaCl dependency, we changed the NaCl concentration from 500 mM at pH 5.0, to ensure stable $T = 3$ capsids, to 100 mM at pH 7.5, to reduce the strength of ELP-interactions and to ensure dynamics. Monitoring the capsid assembly state (Figure 2A,B) shows that after a short lag time, on the same time scale needed for the equilibration of the NaCl concentration during dialysis (Figure S2B), the shift from $T = 3$ to $T = 1$ capsids takes place via a rapid initial process, followed by one that is more gradual (Figure 2B). The complete capsid size transition takes months. Further evaluation with TEM (Figure 2E) confirms this transition process. To get a better understanding of the mechanism of transition, we performed experiments in which we added fluorescently labeled ELP-CPs to unlabeled capsids. We observe that at both pH 5.0 and 7.5 the capsids can exchange dimers with the solution (Figures S11 and S12), which makes it plausible that the observed size change involves the transfer of dimers. Furthermore, we note that it is unlikely for one structure to morph into the other one without disassembly because of the change in the radius of curvature between the two structures. If the sizes of the two structures were close to each other, then it would be possible for the big pieces of one shell to be recycled to form another shell.[28]

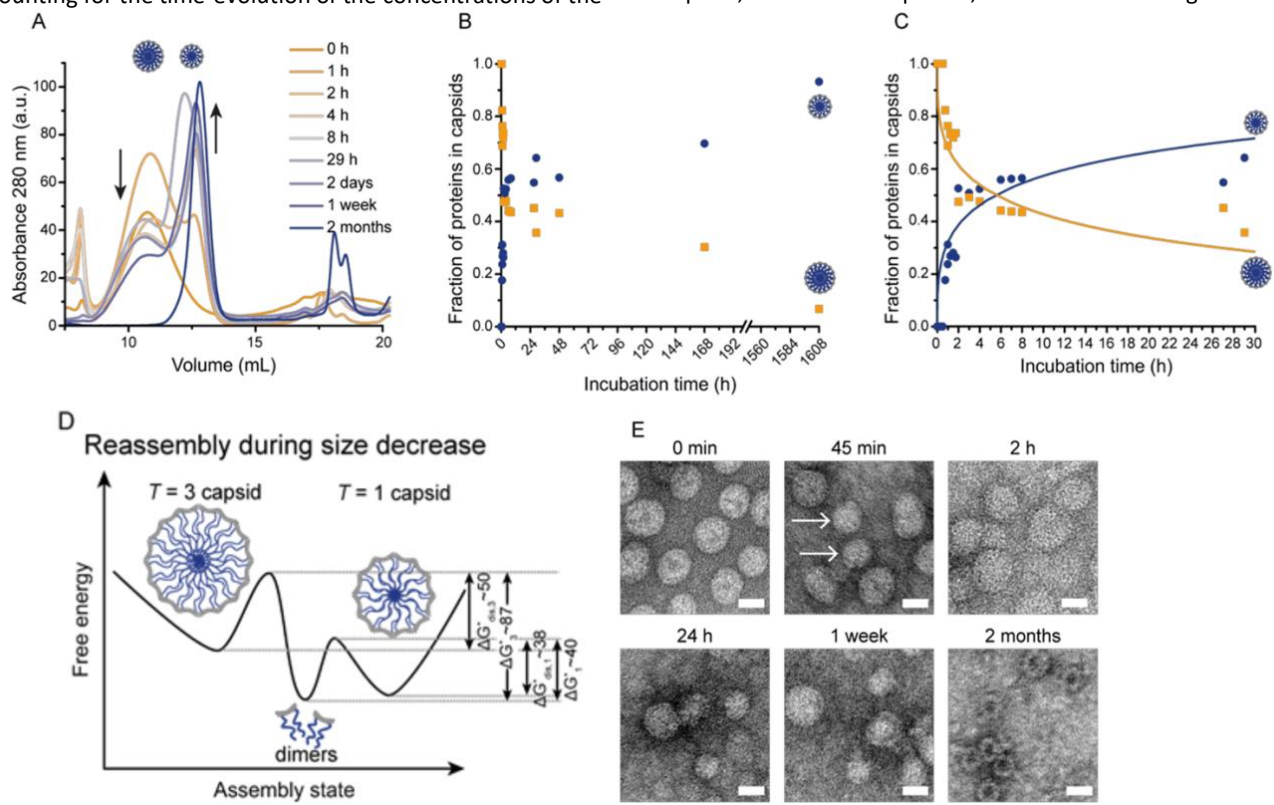

**Figure 2.** Analysis of ELP-CCMV capsids during the transition from $T = 3$ to $T = 1$ particles at pH 7.5. (A) SEC chromatograms measured after indicated dialysis times to pH 7.5. (B and C) Protein fractions as $T = 1$ (blue circles) and $T = 3$ (yellow squares) capsids as determined by integration of the SEC chromatograms (see also Figures S7-S9). The solid lines are the results of our numerical solution (eqs 3 and 4). See Table S4 for more details. (D) Schematic overview of the proposed reassembly mechanism during size decrease, where $T = 1$ capsids are energetically most favorable under the buffer conditions used. ΔG values are in $k_B T$ units. Energy barriers are not drawn to scale; the values provided are indicative. (E) TEM micrographs of samples that were taken after the indicated dialysis times. $T = 1$ capsids in the 45 min image are indicated with arrows. Scale bars correspond to 20 nm. Overview images and additional time points are depicted in Figure S10.

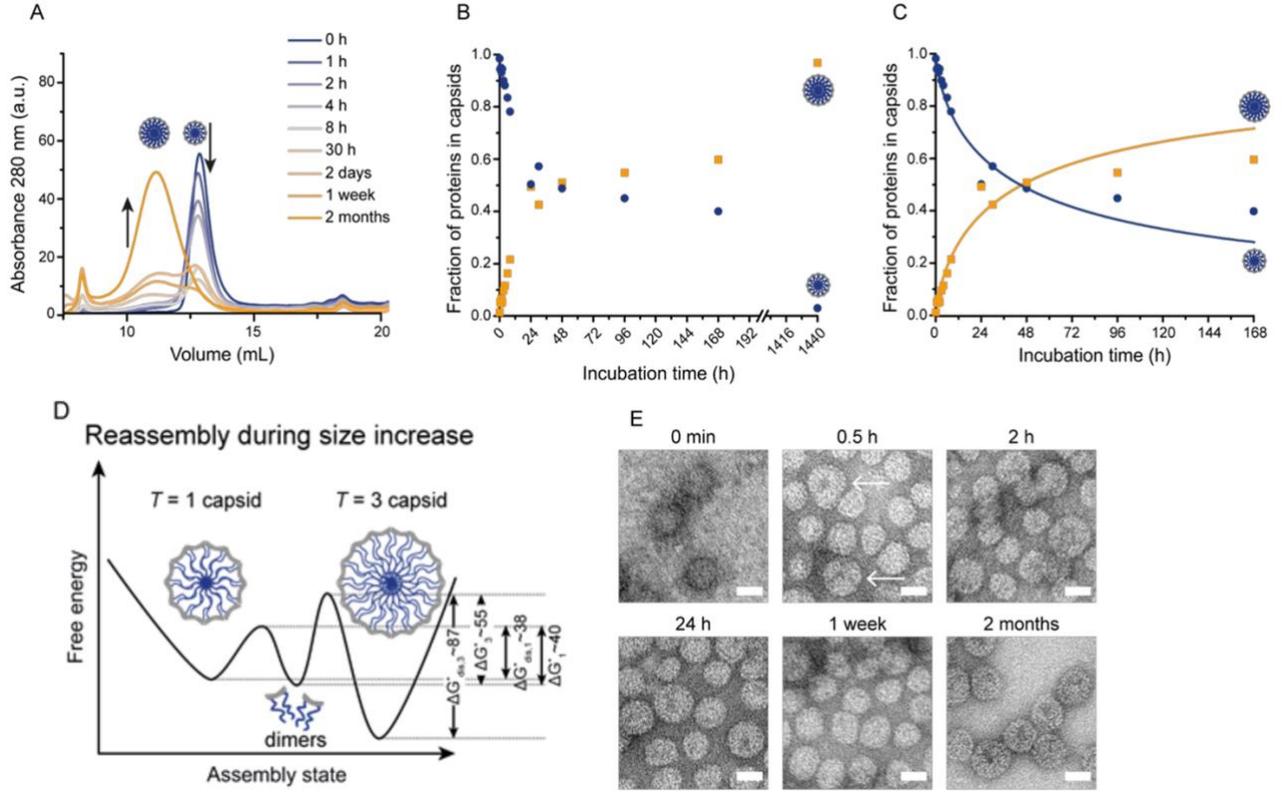

**Figure 3.** Analysis of ELP-CCMV capsids during transition from $T = 1$ to $T = 3$ particles at pH 5.0. (A) SEC chromatograms measured after indicated dialysis times to pH 5.0. (B and C) Protein fractions as $T = 1$ (blue circles) and $T = 3$ (yellow squares) capsids as determined by integration of the SEC chromatograms ( See also Figures S13- S15). The solid lines are the results of our numerical solution (eqs 3 and 4). See Table S5 for more details. (D) Schematic overview of the proposed reassembly mechanism during size increase, where $T = 3$ capsids are energetically most favorable under the buffer conditions used. ∆G values are in $k_B T$ units. Energy barriers are not drawn to scale; the values provided are indicative. (E) TEM micrographs of samples taken after the indicated dialysis times. The $T = 3$ capsids in the 0.5 h image are indicated with arrows. Scale bars correspond to 20 nm. Overview images and additional time points are depicted in Figure S16.

Our experiments suggest that we are pitting the assembly rate of one species against the disassembly rate of another. In order to explore the role of metastability in our experiments, we resort to CNT, as a plausible model to describe the system. Within CNT, the steady-state capsid assembly and disassembly rates $J_{as,T}$ and $J_{dis,T}$ can be written as [26,29]

$$J_{as,T} = x_s \, v^*_T Z_T \, exp \, (-\Delta G^*_{as,T}), \quad (1)$$

$$J_{dis,T} = x_T \, v^*_T Z_T \, exp \, (-\Delta G^*_{dis,T}). \quad (2)$$

Where, $v_T^*$, $Z_T$, $x_s$ and $x_T$ denote the attempt frequency of dimers attaching to the critical nucleus, the Zeldovich factor, and the mole fraction of free subunits, and the capsid of a given T number (Supporting Information section 3.1). $\Delta G^*_{as,T}$ is the height of energy barrier between the free proteins and fully formed capsids, while $\Delta G^*_{dis,T}$ is the height of the free energy barrier between the assembled and free CPs (Figures 2D, and 3D for the opposite size shift). The barrier height depends on the overall protein concentration and on the binding free energies of the proteins in the two types of shell, $g_T$, in units of thermal energy, averaged over all subunits of a fully formed capsid. The kinetic equations describing the concentration of dimers and $T = 1$ and $T = 3$ capsids read as

$$\frac{dx_s}{dt} = -q_1 J_{as,1} - q_3 J_{as,3} + q_1 J_{dis,1} + q_3 J_{dis,3}, \quad (3)$$

and

$$\frac{dx_T}{dt} = J_{as,T} - J_{dis,T}. \quad (4)$$

where $q_1$ and $q_3$ are the numbers of dimers in fully formed capsids of $T = 1$ and 3, respectively. The quantities on the left-hand sides of eqs 3 and 4 represent time derivatives of the concentrations of the species in our model. The terms on the right-hand sides are due to the formation or dissociation of capsids. We solve the above system of equations numerically, using an explicit forward Euler method with adaptive time steps (Supporting Information section 3.2).

Consistent with the experiments (Figure 2B), we find that upon increasing the pH from 5 to 7.5, the amount of $T = 3$ structures decreases while at the same time the number of $T = 1$ structures increases, indicating that under these experimental conditions the protein–protein attraction is stronger between subunits forming $T = 1$ shells than that of those forming $T = 3$ ones. Our curve fits in Figure 2C for times up to 30 h give $g_1 = -15.0$ and $g_3 = -14.7$ in thermal energy units (Supporting Information section 3.1).

As the $T = 3$ shells disassemble, the concentration of free dimers increases and, at some point, reaches the value of the critical capsid concentration $c^*_1 = e^{g_1}$, whereupon $T = 1$ shells start forming and consuming free dimers. As the free dimer concentration continues to increase, the disassembly rate of the $T = 3$ shells decreases, and the assembly rate of $T = 1$ shells increases, explaining the behavior of the disassembly and assembly curves shown in Figure 2C. However, fairly quickly the free dimer concentration attains a more or less constant value because the disassembly of $T = 3$ shells produces dimers that are immediately depleted by the formation of $T = 1$ shells, confirming that the changes in protein fraction in the capsids are due to the disassembly of $T = 3$ and assembly of $T = 1$ (Supporting Information section 3.3). We note that the decrease in free dimer concentration after two months in Figure 2A could be due to the fact that dimer proteins at pH 7.5 after prolonged storage are not highly stable and some aggregation and denaturation will occur over time. The theory presented in this paper does not include this effect.

We next discuss the size shift from $T = 1$ to $T = 3$ following a jump in pH from 7.5 to 5 at a constant NaCl concentration of 500mM. Herein, a 100 μM solution of empty VW1-VW8 ELP-CCMV $T = 1$ capsids in a pH 7.5 buffer with 500 mM NaCl was dialyzed to a pH 5.0 buffer with 500 mM NaCl at 4 °C, during which the capsid assembly state was monitored with SEC and TEM measurements. Figure 3A,B shows that $T = 1$ particles, stable at neutral pH, disappear over time, while $T = 3$ particles appear. The whole process proceeds much more gradually than the opposite size shift and takes around 2 months to reach full completion (Figure 3B). We follow the dynamics with TEM (Figure 3E), confirming the increase in the number of $T = 3$ particles.

The number of $T = 1$ structures decreases and the amount of $T = 3$ structures increases in parallel, which points at stronger attractive interactions between CPs in the native species at low pH. Our curve fits in Figure 3C for times up to 168 h give $g_1 = -15.0$ and $g_3 = -15.4$ in thermal energy units. Again we find that the free subunit concentration very quickly becomes more or less constant: The disassembly of $T = 1$ shells produces dimers that are used for the formation of $T = 3$ shells.

From Figures 2B,C and 3B,C, it appears that $T = 3$ capsids easily dissociate at pH 7.5, crossing the growing fraction of $T = 1$ capsids after 6 h, while the disassembly of $T = 1$ CPs at pH 5.0 is much slower, crossing the growing fraction of $T = 3$ capsids only after 48 h. This is expected because the smaller size of a $T = 1$ capsid produces fewer subunits per disassembled shell. ELPs are positioned closer next to each other because of the higher curvature of $T = 1$ shells, and the interaction between ELPs remains strong at pH 5.0.

In this context we note that under certain conditions the association and dissociation of empty capsids is characterized by hysteresis: it is easier for capsids to assemble than to disassemble.[30] Hence, assembled capsids can be significantly more stable kinetically than they are thermodynamically, implying that the height of the free energy barrier must be larger for disassembly than it is for assembly.[26,31] For the experiments described in this paper, this means that the disassembly step must be rate-limiting if the unstable shells are to be converted into stable shells of a different size. This is indeed what we also find from our theoretical calculations.

In conclusion, we find that ELP-CPs can reversibly switch between $T = 1$ and 3 structures upon changing the solution conditions. While we have not ruled out the possibility that other models can also describe our experiments, remarkably, the interconversion between the two structures can be quite accurately described at least for initial and intermediate times by CNT. At pH 7.5, the driving force for the assembly of coat proteins is the interaction between the ELPs, while at pH 5.0 the attractive interaction between capsid proteins predominates over the attractive ELP−ELP interactions. Since ELPs are attached to the capsid proteins, the ELP-CCMVs do form a shell at pH 7.5, but only the smallest possible one as the ELPs need to be as close as possible to each other to make contact. This insight is of importance not only for a more fundamental understanding of virus assembly but also for the improved design of VLP-based nanomedicines.

## ASSOCIATED CONTENT

### Supporting Information

Experimental methods for protein expression and analysis of self-assembly dynamics, supplemental discussions regarding the method optimizations, theoretical methods and discussions, and additional SEC chromatograms and TEM micrographs.

## AUTHOR INFORMATION

### Corresponding Author

**Jan C. M. van Hest** − *Bio-Organic Chemistry Research Group, Institute for Complex Molecular Systems, Eindhoven University of Technology, 5600 MB Eindhoven, The Netherlands;* orcid.org/0000-0001-7973-2404;

Email: j.c.m.v.hest@tue.nl

**Roya** Zandi − *Department of Physics and Astronomy, University of California, Riverside, California 92521, United States; ;* orcid.org/0000-0001-8769-0419

Email: royaz@ucr.edu

Authors

**Suzanne B. P. E.** Timmermans – *Bio-Organic Chemistry Research Group, Institute for Complex Molecular Systems, Eindhoven University of Technology, 5600 MB Eindhoven, The Netherlands; orcid.org/0000-0003-1073-9004*

**Alireza Ramezani** – *Department of Physics and Astronomy, University of California, Riverside, California 92521, United States; orcid.org/0000-0003-0570-9935*

**Toni Montalvo** – *Department of Physics and Astronomy, University of California, Riverside, California 92521, United States*

**Mark Nguyen** – *Department of Physics and Astronomy, University of California, Riverside, California 92521, United States*

**Paul van der Schoot** – *Soft Matter and Biological Physics Group, Department of Applied Physics, Eindhoven University of Technology, 5600 MB Eindhoven, The Netherlands*

Author Contributions

‖ S.B.P.E.T. and A.R. contributed equally to this work.
The manuscript was written through contributions of all authors. All authors have given approval to the final version of the manuscript.

Notes

The authors declare no competing financial interest.

ACKNOWLEDGMENT

S.B.P.E.T. and J.C.M.v.H acknowledge financial support from the European Research Council (ERC Advanced Grant Artisym 694120) and the Dutch Ministry of Education, Culture and Science (Gravitation program 024.001.035). RZ acknowledges support from NSF DMR-2131963 and the University of California Multicampus Research Programs and Initiatives (grant No. M21PR3267).

ABBREVIATIONS

BMV, Brome mosaic virus; CCMV, cowpea chlorotic mottle virus; CNT, classical nucleation theory; CP, coat protein; ELP, elastin-like polypeptide; ss, single-stranded; SEC, size-exclusion chromatography; TEM, transmission electron microscopy; VLP, viruslike particle.

REFERENCES

(1) Bruinsma, R. F.; Wuite, G. J. L.; Roos, W. H. Physics of Viral Dynamics. Nat. Rev. Phys. 2021, 1–16. https://doi.org/10.1038/s42254-020-00267-1.

(2) Hagan, M. F. Modeling Viral Capsid Assembly. Adv. Chem. Phys. 2014, 155, 1–68. https://doi.org/10.1002/9781118755815.ch01.

(3) Bancroft, J. B.; Hiebert, E. Formation of an Infectious Nucleoprotein from Protein and Nucleic Acid Isolated from a Small Spherical Virus. Virology 1967, 32 (2), 354–356. https://doi.org/10.1016/0042-6822(67)90284-x.

(4) Bancroft, J. B. The Self-Assembly of Spherical Plant Viruses. Adv. Virus Res. 1970, 16, 99–134. https://doi.org/10.1016/s0065-3527(08)60022-6.

(5) Adolph, K. W.; Butler, P. J. Studies on the Assembly of a Spherical Plant Virus. I. States of Aggregation of the Isolated Protein. J. Mol. Biol. 1974, 88 (2), 327–341. https://doi.org/10.1016/0022-2836(74)90485-9.

(6) Chevreuil, M.; Lecoq, L.; Wang, S.; Gargowitsch, L.; Nhiri, N.; Jacquet, E.; Zinn, T.; Fieulaine, S.; Bressanelli, S.; Tresset, G. Nonsymmetrical Dynamics of the HBV Capsid Assembly and Disassembly Evidenced by Their Transient Species. J. Phys. Chem. B 2020, 124 (45), 9987–9995. https://doi.org/10.1021/acs.jpcb.0c05024.

(7) Zhou, J.; Zlotnick, A.; Jacobson, S. C. Disassembly of Single Virus Capsids Monitored in Real Time with Multicycle Resistive-Pulse Sensing. Anal. Chem. 2022, 94 (2), 985–992. https://doi.org/10.1021/acs.analchem.1c03855.

(8) Yamauchi, Y.; Greber, U. F. Principles of Virus Uncoating: Cues and the Snooker Ball. Traffic Cph. Den. 2016, 17 (6), 569–592. https://doi.org/10.1111/tra.12387.

(9) Michaels, T. C. T.; Bellaiche, M. M. J.; Hagan, M. F.; Knowles, T. P. J. Kinetic Constraints on Self-Assembly into Closed Supramolecular Structures. Sci. Rep. 2017, 7 (1), 12295. https://doi.org/10.1038/s41598-017-12528-8.

(10) Hagan, M. F.; Chandler, D. Dynamic Pathways for Viral Capsid Assembly. Biophys. J. 2006, 91 (1), 42–54. https://doi.org/10.1529/biophysj.105.076851.

(11) Keef, T.; Micheletti, C.; Twarock, R. Master Equation Approach to the Assembly of Viral Capsids. J. Theor. Biol. 2006, 242 (3), 713–721. https://doi.org/10.1016/j.jtbi.2006.04.023.

(12) Harms, Z. D.; Selzer, L.; Zlotnick, A.; Jacobson, S. C. Monitoring Assembly of Virus Capsids with Nanofluidic Devices. ACS Nano 2015, 9 (9), 9087–9096. https://doi.org/10.1021/acsnano.5b03231.

(13) Moerman, P.; van der Schoot, P.; Kegel, W. Kinetics versus Thermodynamics in Virus Capsid Polymorphism. J. Phys. Chem. B 2016, 120 (26), 6003–6009. https://doi.org/10.1021/acs.jpcb.6b01953.

(14) Zlotnick, A. To Build a Virus Capsid: An Equilibrium Model of the Self Assembly of Polyhedral Protein Complexes. J. Mol. Biol. 1994, 241 (1), 59–67. https://doi.org/10.1006/jmbi.1994.1473.

(15) Rapaport, D. C. Role of Reversibility in Viral Capsid Growth: A Paradigm for Self-Assembly. Phys. Rev. Lett. 2008, 101 (18), 186101. https://doi.org/10.1103/PhysRevLett.101.186101.

(16) Zandi, R.; van der Schoot, P. Size Regulation of Ss-RNA Viruses. Biophys. J. 2009, 96 (1), 9–20. https://doi.org/10.1529/biophysj.108.137489.

(17) Garmann, R. F.; Comas-Garcia, M.; Knobler, C. M.; Gelbart, W. M. Physical Principles in the Self-Assembly of a Simple Spherical Virus. Acc. Chem. Res. 2016, 49 (1), 48–55. https://doi.org/10.1021/acs.accounts.5b00350.

(18) Zandi, R.; Dragnea, B.; Travesset, A.; Podgornik, R. On Virus Growth and Form. Phys. Rep. 2020, 847 (C). https://doi.org/10.1016/j.physrep.2019.12.005.

(19) Sun, X.; Cui, Z. Virus-Like Particles as Theranostic Platforms. Adv. Ther. 2020, 3 (5), 1900194. https://doi.org/10.1002/adtp.201900194.

(20) van Eldijk, M. B.; Wang, J. C.-Y.; Minten, I. J.; Li, C.; Zlotnick, A.; Nolte, R. J. M.; Cornelissen, J. J. L. M.; van Hest, J. C. M. Designing Two Self-Assembly Mechanisms into One Viral Capsid Protein. J. Am. Chem. Soc. 2012, 134 (45), 18506–18509. https://doi.org/10.1021/ja308132z.

(21) Urry, D. W. Physical Chemistry of Biological Free Energy Transduction As Demonstrated by Elastic Protein-Based Polymers. J. Phys. Chem. B 1997, 101 (51), 11007–11028. https://doi.org/10.1021/jp972167t.

(22) Timmermans, S. B. P. E.; Vervoort, D. F. M.; Schoonen, L.; Nolte, R. J. M.; van Hest, J. C. M. Self-Assembly and Stabilization of Hybrid Cowpea Chlorotic Mottle Virus Particles under Nearly Physiological Conditions. Chem. Asian J. 2018, 13 (22), 3518–3525. https://doi.org/10.1002/asia.201800842.

(23) Schoonen, L.; Maas, R. J. M.; Nolte, R. J. M.; van Hest, J. C. M. Expansion of the Assembly of Cowpea Chlorotic Mottle Virus towards Non-Native and Physiological Conditions. Honor Profr. Ben Feringa 2016 Tetrahedron Prize Creat. Org. Chem. Dyn. Funct. Mol. Syst. 2017, 73 (33), 4968–4971. https://doi.org/10.1016/j.tet.2017.04.038.

(24) Panahandeh, S.; Li, S.; Marichal, L.; Leite Rubim, R.; Tresset, G.; Zandi, R. How a Virus Circumvents Energy Barriers to Form Symmetric Shells. ACS Nano 2020, 14 (3), 3170–3180. https://doi.org/10.1021/acsnano.9b08354.

(25) Panahandeh, S.; Li, S.; Zandi, R. The Equilibrium Structure of Self-Assembled Protein Nano-Cages. Nanoscale 2018, 10 (48), 22802–22809. https://doi.org/10.1039/c8nr07202g.

(26) Zandi, R.; Schoot, P. van der; Reguera, D.; Kegel, W.; Reiss, H. Classical Nucleation Theory of Virus Capsids. Biophys. J. 2006, 90 (6), 1939–1948. https://doi.org/10.1529/biophysj.105.072975.

(27) Kashchiev, D. In Nucleation; Elsevier, 2000; pp 113–290.

(28) Panahandeh, S.; Li, S.; Dragnea, B.; Zandi, R. Virus Assembly Pathways Inside a Host Cell. *ACS Nano* **2022**, *16* (1), 317–327. https://doi.org/10.1021/acsnano.1c06335.

(29) Luque Santolaria, A. *Structure, Mechanical Properties, and Self-Assembly of Viral Capsids*; Universitat de Barcelona, 2011.

(30) Singh, S.; Zlotnick, A. Observed Hysteresis of Virus Capsid Disassembly Is Implicit in Kinetic Models of Assembly. *J. Biol. Chem.* **2003**, *278* (20), 18249–18255. https://doi.org/10.1074/jbc.M211408200.

(31) van der Schoot, P.; Zandi, R.

Supporting Information

# The dynamics of virus-like capsid assembly and disassembly

Suzanne B. P. E. Timmermans,[1,‖] Alireza Ramezani,[2,‖] Toni Montalvo,[2] Mark Nguyen,[2] Paul van der Schoot,[3] Jan C. M. van Hest,[1*] Roya Zandi[2*]

1 Bio-Organic Chemistry Research Group, Institute for Complex Molecular Systems, Eindhoven University of Technology, P.O. Box 513, 5600 MB Eindhoven, The Netherlands; 2 Department of Physics and Astronomy, University of California Riverside, California 92521, USA; 3 Soft Matter and Biological Physics Group, Department of Applied Physics, Eindhoven University of Technology, P.O. Box 513, 5600 MB Eindhoven, The Netherlands.

[‖]   These authors contributed equally to this work.

# Contents

# 1 Experimental materials and methods

## 1.1 Materials

Ampicillin, chloramphenicol, yeast extract, and peptone were purchased from Sigma-Aldrich/ Merck. Isopropyl-β-D-thiogalactopyranoside (IPTG) was obtained from PanReac AppliChem VWR. Ni-NTA agarose beads were obtained from Qiagen.

## 1.2 Buffers

**Table S1 – Composition of buffers**

| Name | Composition |
|---|---|
| pH 5.0 buffer | 50 mM NaOAc, 500 mM NaCl, 10 mM $MgCl_2$, 1 mM EDTA, pH 5.0 |
| pH 5.0 no EDTA buffer | 50 mM NaOAc, 500 mM NaCl, 10 mM $MgCl_2$, pH 5.0 |
| pH 7.5 100 mM NaCl buffer | 50 mM Tris·HCl, 100 mM NaCl, 10 mM $MgCl_2$, 1 mM EDTA, pH 7.5 |
| pH 7.5 100 mM NaCl no EDTA buffer | 50 mM Tris·HCl, 100 mM NaCl, 10 mM $MgCl_2$, pH 7.5 |
| pH 7.5 500 mM NaCl buffer | 50 mM Tris·HCl, 500 mM NaCl, 10 mM $MgCl_2$, 1 mM EDTA, pH 7.5 |
| 4x native PAGE loading buffer | 248 mM Tris·HCl, 40 % glycerol (*v/v*), 0.02 % bromophenol blue (*w/v*) |
| Native PAGE running buffer | 25 mM Tris·HCl, 192 mM glycine |

All buffers were filtered over a 0.2 micron filter prior to use.

## 1.3 UV-vis absorbance measurements

In order to determine the protein concentrations during experiments, the absorbance at 280 nm was measured with a spectrophotometer ND-1000 and the concentrations were subsequently calculated using the theoretical extinction coefficients.[1]

## 1.4 Mass spectrometry

Protein mass characterization was performed using a High Resolution LC-MS system consisting of a Waters ACQUITY UPLC I-Class system coupled to a Xevo G2 Quadrupole Time of Flight (Q-ToF). The system consisted of a Binary Solvent Manager and a Sample Manager with Fixed-Loop (SM-FL). Proteins were separated (0.3 mL/min) on the column (Polaris C18A reverse phase column, 2.0 x 100 mm, Agilent) using an acetonitrile gradient in water (15% to 75%, *v/v*) supplemented with formic acid (0.1%, *v/v*) before analysis in positive mode in the mass spectrometer. Deconvoluted mass spectra were obtained using the MaxENT1 algorithm in the Masslynx v4.1 (SCN862) software. Isotopically averaged molecular weights were calculated using the 'Protein Calculator v3.4' at http://protcalc.sourceforge.net. Protein samples were desalted by spin-filtration with MilliQ prior to measurement.

## 1.5  Size exclusion chromatography (SEC)

SEC analysis was performed on a Superose 6 increase 10/300 GL column (GE Healthcare Life Sciences). Analytical measurements were executed on an Agilent 1260 bio-inert HPLC. Samples with a protein concentration of 100 µM were separated on the column at 21 °C with a flow rate of 0.5 mL/min. Running buffer was "pH 7.5, 100 mM NaCl no EDTA buffer" for $T = 3$ to $T = 1$ shift and "pH 5.0 buffer" for $T = 1$ to $T = 3$ shift (see table S1 for exact compositions).

## 1.6  Transmission electron microscopy (TEM)

TEM grids (FCF-200-Cu, EMS) were glow-discharged using a Cressington 206 carbon coater and power unit. Protein samples (10 µM, 5 µL) were applied on the glow-discharged grids and incubated for 1 min. The samples were carefully removed using a filter paper. Then, the grid was negatively stained by applying 2% uranyl acetate in water (5 µL). The staining solution was removed after 15 seconds and the grid was allowed to dry for at least 15 minutes. The samples were studied on a FEI Tecnai 20 (type Sphera) (operated at 200 kV, equipped with a $LaB_6$ filament and a FEI BM-Ceta CCD camera).

## 1.7  Dynamic light scattering (DLS) measurements

DLS measurements were performed on a Malvern Zetasizer Nano ZSP at 21°C, unless stated otherwise. Samples (100 µM, unless stated otherwise) were centrifuged twice prior to analysis. Buffers were filtered prior to use. All measurements were done in triplo, and the average of the triplo measurements was plotted.

## 1.8  Conductivity measurements

Conductivity measurements were performed using a Mettler Toledo SevenGo Duo Pro pH/ conductivity meter SG78 that was calibrated with a 1413 µS/cm @ 25 °C standard (VWR, 84135.260). In order to assess the NaCl concentration during dialysis, a standard curve was prepared by measuring the conductivity of a series of pH 7.5 buffers with NaCl concentrations ranging from 0 M to 1 M (with 100 mM intervals). As such, the following formula was used: $C(mS/cm) = 0.076 \times [NaCl](mM) + 6.1536$ with $R^2 = 0.9991$.

## 1.9 Native PAGE

Samples were prepared by mixing 7.5 µL of a 100 µM protein solution and 2.5 µL of 4x native PAGE loading buffer. 2.67 µL of the samples were loaded on a 4-20 % gel (Bio-Rad, Mini-PROTEAN TGX Precast Gels cat# 456-1096). The gel was run at 4 °C with pre-cooled native PAGE running buffer containing a cooling pack for 3 hours at 100 V. The gel was washed for 5 minutes in demineralized water and then stained with coomassie blue staining solution (Bio-Rad, cat# 161-0786). Subsequently, the gel was destained in demineralized water and visualized with an ImageQuant 350 gel imager with ImageQuant 350 Capture software.

## 1.10 General protocol for the expression of $His_6$-ELP-CCMV variants

The pET-15b-$His_6$-ELP-CCMV(ΔN26), pET-15b-$His_6$-VW1-VW8 ELP-CCMV(ΔN26) and pET15b-mEGFP-$His_6$-VW1-VW8 ELP-CCMV(ΔN26) vectors encoding for the three ELP-CCMV variants used in this manuscript were previously constructed.[2,3]

The native ELP-CCMV, VW1-VW8 ELP-CCMV and mEGFP-VW1-VW8 ELP-CCMV capsid protein variants were expressed according to standard expression procedures.[2,4,5] As an example, *E.coli* BLR(DE3)pLysS glycerol stocks containing the pET-15b-$H_6$-VW1-VW8-ELP-CCMV(ΔN26) vector were used for an overnight culture at 37 °C in LB medium (50 mL), containing ampicillin (100 mg/L) and chloramphenicol (50 mg/L). The overnight culture was used to inoculate 2xTY medium (1 L), containing ampicillin (100 mg/L) and grown at 37 °C till an optical density ($OD_{600}$) between 0.4 and 0.6 was reached. Protein expression was then induced by addition of IPTG (1 mM) and the culture was incubated at 30°C for 6 hours. Cells were harvested by centrifugation (2700 x *g*, 15 min, and 4 °C) and pellets were stored overnight at -20 °C.

The cell pellet was resuspended in lysis buffer (50 mM $NaH_2PO_4$, 300 mM NaCl, 10 mM imidazole, pH 8.0; 25 mL). Cell lysis was performed by ultrasonic disruption (7-10 times 30 seconds, 70% amplitude, Branson Sonifier 150). Cell lysate was then centrifuged (16.000 x *g*, 15 min, 4 °C) to remove bacterial debris. The supernatant was incubated with Ni-NTA agarose beads (3 mL) for 1 hour at 4 °C, followed by column loading. The flow-through was collected and the column was washed twice with wash buffer (50 mM $NaH_2PO_4$, 300 mM NaCl, 20 mM imidazole, pH 8.0; 20 mL). The proteins of interest were eluted from the column with elution buffer (50 mM $NaH_2PO_4$, 300 mM NaCl, 250 mM imidazole, pH 8.0; 1 time 0.5 mL, 7 times 1.5 mL) and fractions containing histidine-tagged VW1-VW8 ELP-CCMV were combined and dialyzed against pH 7.5 dimer buffer (50 mM Trizma base, 500 mM NaCl, 10 mM MgCl$_2$, 1 mM EDTA, pH 7.5; 3 times 30-60 minutes using a 12-14 kDa tubing). The protein was then dialyzed against pH 5.0 capsid buffer (50 mM Trizma base, 500 mM NaCl, 10 mM MgCl$_2$, 1 mM EDTA, pH 5.0; 2 times 30-60 minutes, followed by overnight dialysis using a 12-14 kDa tubing) for stable storage at 4°C. The purity and characteristics of the protein were verified and determined by SDS-PAGE, SEC, Q-TOF mass spectrometry, DLS and TEM. The protein yields after purification and the Q-TOF results are listed in Table S2. The amino acid sequences of the three ELP-CCMV variants are listed in Table S3.

**Table S2 – Expression yields and Q-TOF results of ELP-CCMV variants**

| Name | Yield (mg/L) | Q-TOF results | |
|---|---|---|---|
| | | Calculated MW (Da) | Observed MW (Da) |
| Native ELP-CCMV | 10-48 | 22253.4 | 22253.4 |
| VW1-VW8 ELP-CCMV | 22-34 | 22427.6 | 22427.2 |
| mEGFP-VW1-VW8 ELP-CCMV | 15-22 | 49807.5 | 49807.4 |

**Table S3 – Amino Acid sequences of ELP-CCMV variants**

| Name | Sequence |
|---|---|
| Native ELP-CCMV | GHHHHHHVPGVGVPGLGVPGVGVPGLGVPGVGVPGLGVPGGGVPGVGVPGLGLEVVQPVIVEPIASGQGKAIKAWTGYSVSKWTASCAAAEAKVTSAITISLPNELSSERNKQLKVGRVLLWLGLLPSVSGTVKSCVTETQTTAAASFQVALAVADNSKDVVAAMYPEAFKGITLEQLTADLTIYLYSSAALTEGDVIVHLEVEHVRPTFDDSFTPVY |
| VW1-VW8 ELP-CCMV | GHHHHHHVPG**W**GVPGLGVPGVGVPGLGVPGVGVPGLGVPGGGVPG**W**GVPGLGLEVVQPVIVEPIASGQGKAIKAWTGYSVSKWTASCAAAEAKVTSAITISLPNELSSERNKQLKVGRVLLWLGLLPSVSGTVKSCVTETQTTAAASFQVALAVADNSKDVVAAMYPEAFKGITLEQLTADLTIYLYSSAALTEGDVIVHLEVEHVRPTFDDSFTPVY |
| mEGFP-VW1-VW8 ELP-CCMV | MVSKGEELFTGVVPILVELDGDVNGHKFSVSGEGEGDATYGKLTLKFICTTGKLPVPWPTLVTTLTYGVQCFSRYPDHMKQHDFFKSAMPEGYVQERTIFFKDDGNYKTRAEVKFEGDTLVNRIELKGIDFKEDGNILGHKLEYNYNSHNVYIMADKQKNGIKVNFKIRHNIEDGSVQLADHYQQNTPIGDGPVLLPDNHYLSTQSKLSKDPNEKRDHMVLLEFVTAAGITLGMDELYKGSGSMGHHHHHHVPG**W**GVPGLGVPGVGVPGLGVPGVGVPGLGVPGGGVPG**W**GVPGLGLEVVQPVIVEPIASGQGKAIKAWTGYSVSKWTASCAAAEAKVTSAITISLPNELSSERNKQLKVGRVLLWLGLLPSVSGTVKSCVTETQTTAAASFQVALAVADNSKDVVAAMYPEAFKGITLEQLTADLTIYLYSSAALTEGDVIVHLEVEHVRPTFDDSFTPVY |

## 1.11 General protocol for measuring the self-assembly dynamics during size reduction (conversion dynamics from T=3 to T=1 particles)

For a typical dynamics experiment a 100 µM VW1-VW8 ELP-CCMV coat protein solution (150 µL – 1200 µL) in pH 5.0 buffer was prepared and dialyzed to "pH 5.0 no EDTA" buffer at 4 °C o.n. (12-14 kDa MWCO). Dialysis buffer (150 mL – 200 mL) was changed after 30 minutes and 60 minutes. Subsequently, the protein solution was dialyzed to "pH 7.5, 100 mM NaCl no EDTA" buffer at 4 °C (12-14 kDa MWCO) with dialysis buffer (150 mL – 200 mL) changes after 30 minutes and 60 minutes. At different time points during dialysis, 110 µL samples were retrieved from the dialysis membrane and incubated with 0.2 eq. of $NiCl_2$ (20 µM) for 50 minutes at room temperature. Subsequently, samples were spun down for 1 minute at 13400 rpm and subjected to SEC analysis. In addition, 5 µL samples were retrieved from the dialysis membrane, spun down for 1 minute at 13400 rpm, and subjected to TEM analysis.

## 1.12 General protocol for measuring the self-assembly dynamics during size increase (conversion dynamics from T=1 to T=3 particles)

For a typical dynamics experiment a 100 µM VW1-VW8 ELP-CCMV coat protein solution (150 µL – 1200 µL) in pH 7.5 buffer with 500 mM NaCl was prepared and dialyzed to pH 5.0 buffer at 4 °C (12-14 kDa MWCO). Dialysis buffer (150 mL – 200 mL) was changed after 30 minutes and 60 minutes. At different time points during dialysis, 110 µL and/or 5 µL samples were retrieved from the dialysis membrane, spun down for 1 minute at 13400 rpm, and subjected to SEC analysis and/ or TEM analysis respectively.

## 1.13 General protocol for mEGFP-labeling of capsids by exchange with labeled dimers

For a typical dynamics-based mEGFP-labeling experiment both VW1-VW8 ELP-CCMV and mEGFP-VW1-VW8 ELP-CCMV protein solutions were dialyzed overnight at 4 °C to pH 7.5 buffer with 100 mM NaCl (12-14 kDa MWCO). Subsequently, the protein solutions were mixed in a 4:1 ratio (VW1-VW8 ELP-CCMV : mEGFP-VW1-VW8 ELP-CCMV) and incubated at 4 °C for 4 hours. Hereafter, the mixtures were dialyzed at 4 °C either to pH 5.0 buffer (*T*=3 capsids) or to pH 7.5 buffer, 500 mM NaCl (*T*=1 capsids) and incubated in the final buffer at 4 °C for up to one week.

At intermediate time points, samples were taken, heated to 21 °C with 1 °C/min and subjected to SEC analysis with pH 5.0 buffer or pH 7.5 buffer with 500 mM NaCl as eluent. At the final time point, fractions were collected during a preparative SEC run and the combined capsid fractions were analyzed with SDS-PAGE and TEM. The amount of mEGFP incorporated into the capsids was determined by SDS-PAGE analysis. Gels that were visualized via Coomassie Brilliant Blue staining (Biorad) were analyzed with ImageJ gel analysis software to calculate the loading of capsids with mEGPF. Hereto, the following formula was used:

$$labelling\ efficiency = \frac{gel_{GFP-ELP-CCMV}/mw_{GFP-ELP-CCMV}}{gel_{ELP-CCMV}/mw_{ELP-CCMV} + gel_{GFP-ELP-CCMV}/mw_{GFP-ELP-CCMV}} \quad \text{Eq. (S1)}$$

where gel is the intensity of the protein band on the SDS-PAGE gel as determined by ImageJ analysis; mw is the molecular weight of the protein.

## 2 Experimental supplemental discussions

### 2.1 Optimization of conditions to allow for optimal dynamics

In order to study VW1-VW8 ELP-CCMV capsid size shifts, we first investigated the optimal conditions that would allow for the dynamic behavior of the capsids. In preliminary results we observed a size shift from $T$ = 1 capsids to $T$ = 3 capsids during overnight dialysis from pH 7.5 to pH 5.0 at 4 °C (Figure S1). We, therefore, evaluated with SEC whether the reverse size shift would also take place. We observed that only a partial shift from $T$ = 3 to $T$ = 1 capsids took place during the overnight dialysis to pH 7.5 buffer with 100 mM NaCl (Figure S1A,B first two chromatograms in each panel). We then proceeded by investigating whether a second overnight dialysis to pH 5.0 would induce a re-shift back to $T$ = 3 capsids as well. Interestingly, this was the case only when the dialysis was performed at 4 °C (Figure S1B), while at 21 °C no size shift was observed. This indicates that at 21 °C the capsids are much less dynamic than at 4 °C. Although this seems contra-intuitive at first, this observation can be explained by the interactions between the hydrophobic ELP-domains which are much stronger at 21 °C than at 4 °C. So it is highly likely that these interactions in the capsid interior hamper rearrangements of the CP domains in the capsid shell, which are necessary for a size shift.

Another factor that was thought to influence the capsid dynamics as a result of ELP interactions is the ionic strength of the buffers used. Previously, we used a pH 7.5 buffer with 100 mM NaCl to completely disassemble other ELP-CCMV variants. Therefore, we hypothesized that this ionic strength would also allow for dynamics within capsids of our more hydrophobic VW1-VW8 ELP-CCMV variant. As, ideally, we wanted to study the capsid size shifts while varying as few factors as possible, preferably only the pH, we attempted to store $T = 3$ capsids in pH 5.0 buffer with 100 mM NaCl instead of 500 mM NaCl. However, this, unfortunately, led to the aggregation of the protein already within 16 hours (data not shown). We, therefore, evaluated whether VW1-VW8 ELP-CCMV exhibited dynamic behavior when dialyzed from pH 5.0 buffer with 500 mM NaCl to pH 7.5 buffer with 500 mM NaCl, thus only changing the pH. Unfortunately, only a very small part of the capsids appeared to be shifted in size after 24 hours (Figure S1C) as compared to dialysis to pH 7.5 buffer with 100 mM NaCl (Figure S1D), which can again be explained by hydrophobic interactions between ELP domains hampering capsid dynamics at 500 mM NaCl. We, therefore, decided to use a shift from pH 5.0 with 500 mM NaCl to pH 7.5 with 100 mM NaCl and vice versa to study the dynamics during VW1-VW8 ELP-CCMV capsid size decrease and increase respectively as is discussed further in the main text.

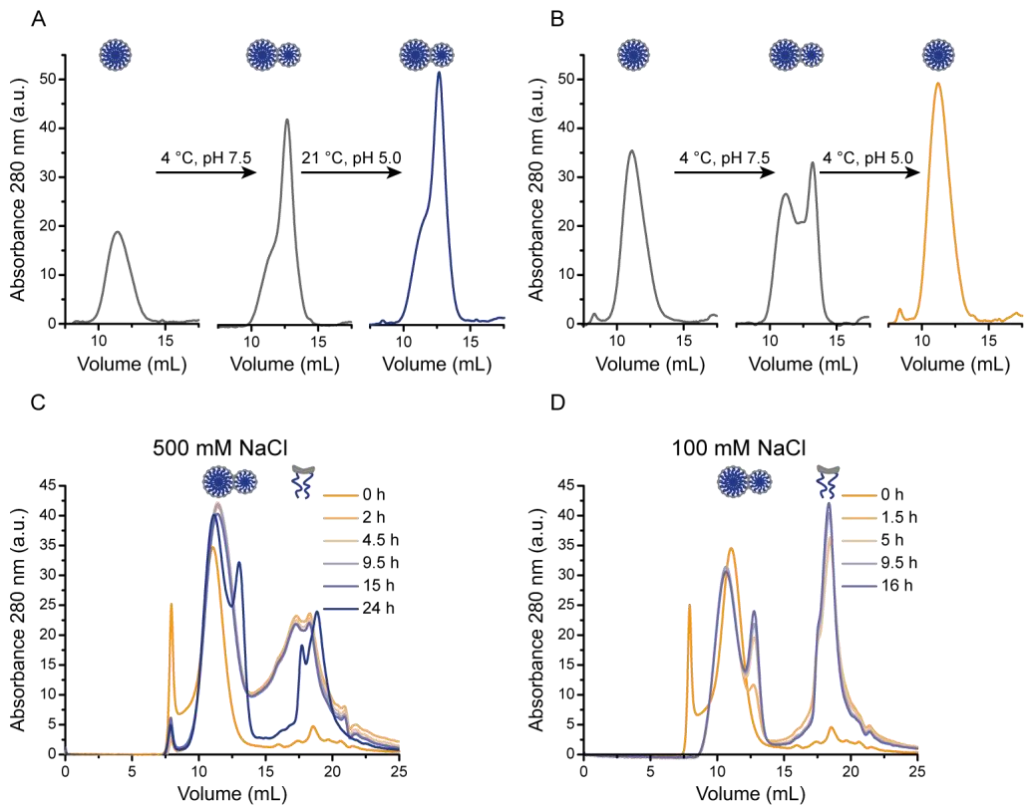

**Figure S1.** Optimization of dynamics during VW1-VW8 ELP-CCMV capsid size changes. A and B) Optimization of temperature. A $T = 3$ capsid solution was dialyzed overnight at 4 °C to pH 7.5 buffer with 100 mM NaCl, yielding a mixture of $T = 3$ and $T = 1$ capsids. Subsequently, the capsid mixture was dialyzed back to pH 5.0 at 21 °C (A) or 4 °C (B), only resulting in a complete shift back to $T = 3$ capsids at the lower temperature. The capsid size was monitored after each dialysis step with SEC via the protein absorbance at 280 nm. C and D) Optimization of NaCl concentration. A $T = 3$ capsid solution was dialyzed overnight at 4 °C to pH 7.5 buffer with 500 mM (C) or 100 mM (D) NaCl. The capsid size was monitored during dialysis with SEC via the protein absorbance at 280 nm. Aggregated proteins or higher order assemblies elute around 7 mL, $T = 3$ capsids around 11 mL, $T = 1$ capsids around 13 mL, and coat protein dimers around 17 mL.

## 2.2 Optimization of dialysis conditions

As described in the previous section, it is necessary to both change the pH and NaCl concentration in order to study size shifts of VW1-VW8 ELP-CCMV capsids. If only the pH would have to be adjusted, this could have been done by adding either HCl or NaOH to the buffer. However, to also adjust the NaCl concentration a buffer exchange step is necessary. Although spin-filtration would be the quickest option and would allow for evaluation of the capsid size upon change of the conditions very quickly, it would also introduce changes in the protein concentration, which could affect the capsid assembly state. As this could complicate our evaluation of capsid size changes as a function of pH, we decided to perform dialysis in order to change the pH and NaCl concentration, despite being a slower process than spin-filtration.

During initial experiments, it was observed that there is a large dependency of capsid dynamics on dialysis time while incubation periods at 4 °C were kept constant (Figure S2A). When a $T = 3$ capsid solution was dialyzed to pH 7.5 buffer with 100 mM NaCl, a large shift to $T = 1$ capsids was only observed during SEC analysis when the total dialysis time was more than 30 minutes. This indicated that either the pH or the NaCl concentration changed slower than anticipated during dialysis. As the pH switch during dialysis takes place within minutes, it was suspected that the other variable during dialysis, the NaCl concentration, changed more slowly. The amount of NaCl that is dissolved in an aqueous solution affects the conductivity of that solution, thus conductivity measurements were performed to follow the change of the NaCl concentration during dialysis. Hereto, a mock dialysis with the same ratio between the volume inside the dialysis bag and the solvent volume was performed and the conductivity was monitored over a time course of 4 hours (Figure S2B). A dialysis time of around 2 hours was necessary to fully convert the NaCl concentration from 500 mM to 100 mM NaCl in the dialysis bag, which explains why dialysis time is such an important determinant of capsid dynamics.

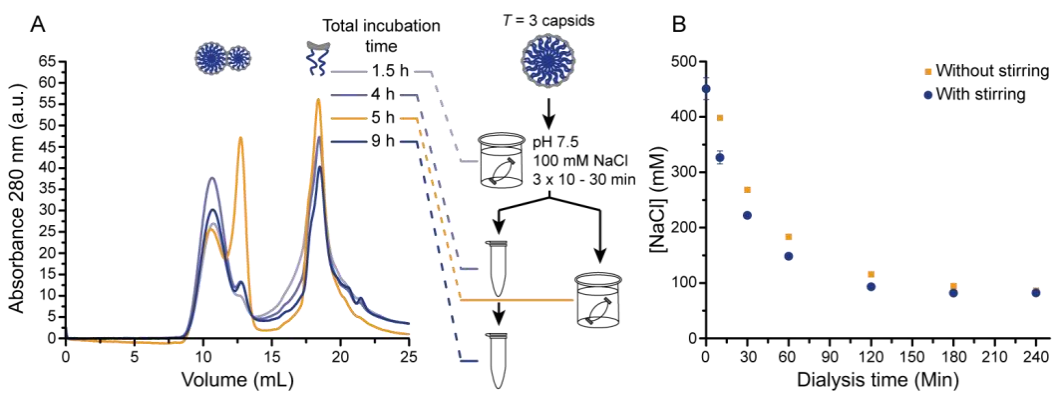

**Figure S2.** Optimization of dialysis conditions. A) SEC chromatograms of samples taken during prolonged dialysis at 4 °C in pH 7.5 buffer with 100 mM NaCl (5 h, orange line) or during incubation in an Eppendorf tube at 4 °C in pH 7.5 buffer with 100 mM NaCl after 3 x 10 minutes initial dialysis (1.5 h, 4 h, 9 h). Thus, while the total incubation time at 4 °C in pH 7.5 buffer with 100 mM NaCl was similar for the 4 h and 5 h samples, the dialysis time was much shorter for the 4 h sample. B) Reduction of the NaCl concentration during dialysis of pH 7.5 buffer with 500 mM NaCl to pH 7.5 buffer with 100 mM NaCl at 4 °C. Buffer was exchanged after 30 and 60 minutes and the dialysis buffer was either stirred at 150 rpm (blue circles) or not stirred (yellow squares). All measurements were performed in triplicate and data is presented as mean ± standard deviation.

## *2.3 Optimization of the SEC protocol for studying the self-assembly dynamics during size reduction*

As during initial dynamics experiments large quantities of dimers were observed in the SEC chromatograms when pH 7.5 buffer with 100 mM NaCl was the eluent, while these were never observed for VW1-VW8 ELP-CCMV before, the origin of these dimers was evaluated. Hereto, a 100 µM VW1-VW8 ELP-CCMV coat protein solution in pH 5.0 buffer was dialyzed (MWCO 12-14 kDa) to pH 7.5 buffer with 100 mM NaCl at 4 °C overnight and subsequently spiked with known amounts of native ELP-CCMV dimers in the same buffer. DLS and native PAGE were employed to analyze the capsid-dimer mixtures. From the DLS results in Figure S3A and B, it becomes clear that DLS is not sensitive enough to detect dimers in capsid-dimer mixtures, which could be explained by the high scattering of the capsids overpowering any light scattering caused by the much smaller dimers. Therefore, although no dimers are detected with DLS of the dialyzed capsids, this does not confirm that indeed no dimers are present in this capsid solution. We therefore focused on native PAGE analysis. From the results in Figure S3C, it can be appreciated that capsids and dimers can be easily distinguished from each other. Furthermore, based on the band intensities on the gel it can be stated that in the VW1-VW8 ELP-CCMV capsid solution less than 5 % dimers are present. This indicates that the large fractions of dimers that are observed in the SEC chromatogram are most likely the result of some disassembly of VW1-VW8 ELP-CCMV capsids taking place during SEC analysis, which might be caused by the extreme dilution (240 times) during the chromatographic procedure.

As for data analysis purposes, the SEC chromatograms should provide the best representation of the assembly state of VW1-VW8 ELP-CCMV during dynamics, a protocol was developed for inhibiting capsid dynamics prior to SEC analysis. Hereto, a 100 µM VW1-VW8 ELP-CCMV solution in pH 5.0 buffer was dialyzed (MWCO 12-14 kDa) to pH 7.5 buffer with 100 mM NaCl at 4 °C overnight and subsequently incubated with various amounts of $NiCl_2$ for 50 minutes (the duration of one SEC run) at room temperature.[6] From the results in Figure S4A and B, it can be observed that the addition of at least 0.2 equivalents $Ni^{2+}$ (relative to the amount of VW1-VW8 ELP-CCMV coat protein concentration) successfully reduces the number of dimers that are observed in the SEC chromatograms, without affecting the $T = 3 : T = 1$ ratios. With increasing amounts of $Ni^{2+}$ also some higher-order structures became visible on the SEC chromatograms (around 7 mL), indicating that high amounts of $Ni^{2+}$ alter the protein fractions in the VW1-VW8 ELP-CCMV solution. Therefore, the addition of 0.2 equivalents of $Ni^{2+}$ was found to be most suitable to reduce the number of dimers introduced due to dilution on the SEC column while not affecting the protein fractions. To confirm that the addition of this amount of $Ni^{2+}$ effectively stops any dynamics, a sample that was dialyzed for 30 minutes was subjected to SEC analysis after 50 minutes or 7 hours of incubation with $Ni^{2+}$ (Figure S4C,D). The resulting SEC traces and protein fractions were very similar, indicating that the addition of 0.2 equivalents of $Ni^{2+}$ effectively stops the dynamics and stabilizes the samples for prolonged storage prior to SEC analysis.

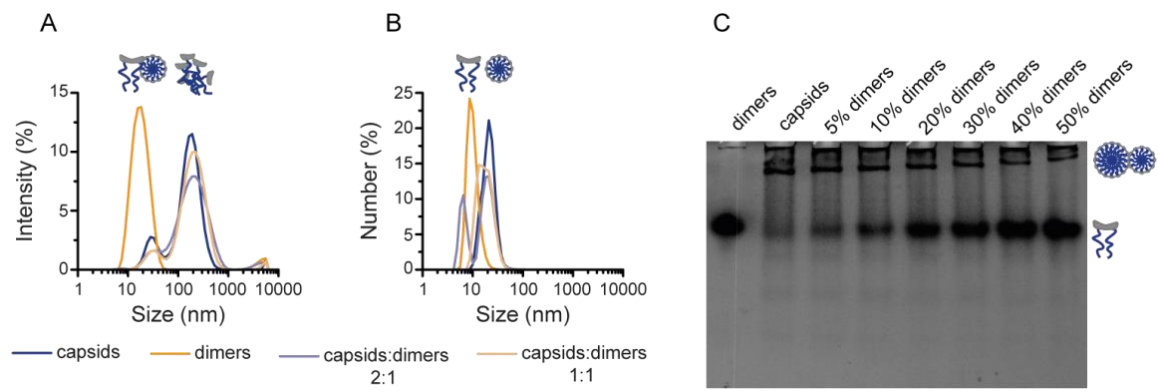

**Figure S3.** Analysis of dimers in capsid mixtures. A and B) DLS analysis of capsids, dimers and capsid-dimer mixtures. Both the intensity (A) and number (B) distributions do not allow for reliable estimation of the number of dimers in the capsid-dimer mixtures. C) Native PAGE analysis of capsids, dimers and capsid-dimer mixtures.

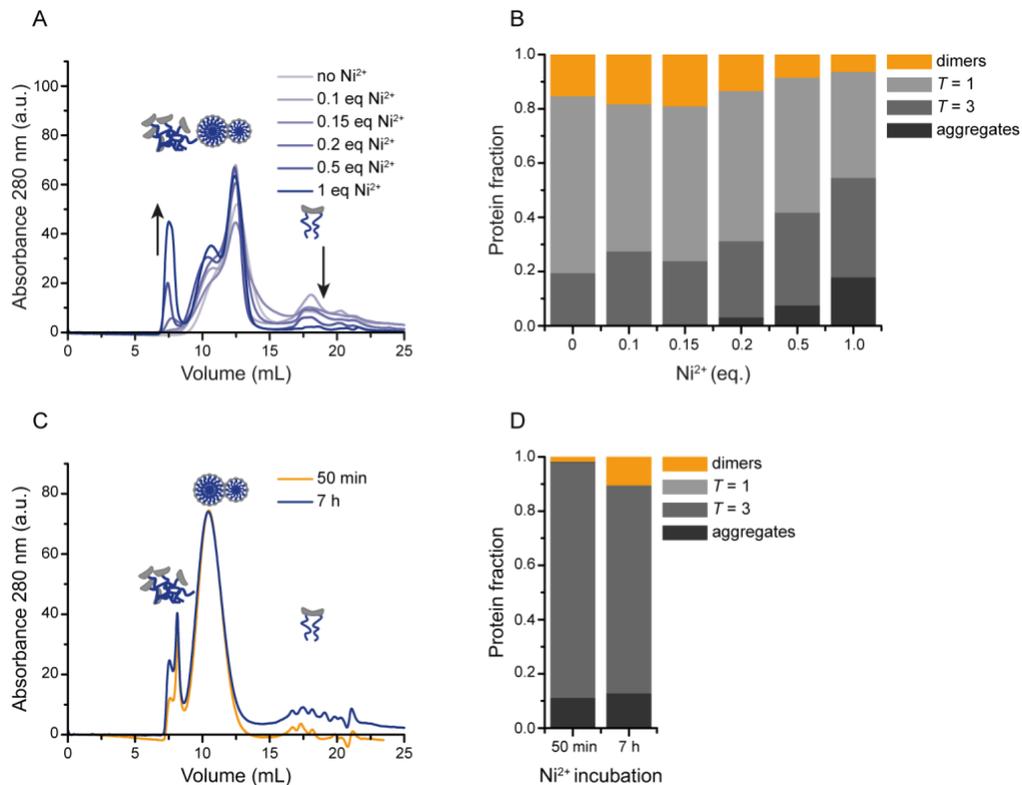

**Figure S4.** Optimization of SEC conditions. A) SEC analysis of capsid mixtures that were stabilized with various amounts of Ni2+. Samples were incubated with the indicated amount of Ni2+ for 50 minutes at room temperature prior to injection. B) Quantification of the protein fraction as dimers, T = 1 capsids, T = 3 capsids and aggregates based on the SEC chromatograms in figure A. C) SEC analysis of capsid mixtures that were stabilized with 0.2 equivalents of Ni2+ for 50 minutes or 7 hours at room temperature prior to injection. D) Quantification of the protein fraction as dimers, T = 1 capsids, T = 3 capsids and aggregates based on the SEC chromatograms in figure C.

# 3 Theoretical methods and discussions

## 3.1 Theory

Based on previous experimental, theoretical and computer simulation results, we put forward that nucleation is the underlying mechanism for capsid assembly and disassembly.[7–9] To this end, we combine equilibrium theory, borrowed from the physics of supramolecular polymers, and classical nucleation theory.[10] This allows us to calculate the time evolution of the assembly and disassembly of mixtures of capsids, the predictions of which we compare with our experimental results. Note that the assembly kinetics of free subunits into competing capsids with different $T$ numbers has been discussed before [11,12] - a kinetic theory of T number conversion has not yet been attempted.

To obtain the thermodynamic parameters required for our kinetic theory, we write the free energy of an aqueous solution in which only free ELP-CCMV subunits and the fully formed capsids are allowed to be present. In the equilibrium theory we ignore the intermediate states, as previous experimental and simulation work, as well as the findings presented in this paper clearly show that they are barely detectable (if at all) and represent short-lived states.[13–17] Within a mean-field approximation, the dimension-less free energy $f$ per solvent molecule of a dilute, aqueous solution, in which the mole fraction of ELP-CP subunits is $x_s$ and that of the fully formed capsids $x_1$ for the $T = 1$ and $x_3$ for $T = 3$ species can be written as

$$f = x_s \ln x_s - x_s + \sum_{T=1,3}[x_T \ln x_T - x_T + q_T g_T x_T] \qquad \text{Eq. (S2)}$$

in units of the thermal energy $k_B T$ with $k_B$ Boltzmann's constant and $T$ the absolute temperature. See Refs.[10,18] for the derivation of Eq. (S2). The effective binding free energy between ELP-CP subunits in $T = 1$ and $T = 3$ capsids are represented by $g_1$ and $g_3$ in units of thermal energy, averaged over all subunits of a fully formed capsid. Finally, $q_1$ and $q_3$ represent the number of subunits that make up the capsid of the $T = 1,3$ shells, respectively. The first four terms in Eq. (S2) represent the translational entropy and the entropy of mixing while the last term accounts for the overall (net) binding free energy of the subunits in assemblies.

An important ingredient in the development of the classical nucleation theory for virus capsids, is the difference between the chemical potential of free protein subunits in the metastable solution and bound proteins in the capsids. The chemical potential of the free protein subunits in solution, $\mu_s = \partial f / \partial x_s$, and chemical potentials of the capsids, $\mu_T = \partial f / \partial x_T$, follow from Eq. (S2) where $T = 1,3$ indicate the $T$ number of the shell. The equilibrium distribution of proteins over the free proteins and different types of capsids can be obtained by minimizing Eq. (S2) subject to the condition of conservation of mass, $c_s = x_s + q_1 x_1 + q_3 x_3$, with $c_s$ the overall mole fraction of coat proteins in solution. The resulting mass-action equations are $x_T = (x_s e^{-g_T})^{q_T}$ with $T = 1,3$. The reference chemical potentials are tacitly absorbed in the binding free energies $g_T$. Since $x_1$ and $x_3$ are always (much) smaller than unity, both $x_s e^{-g_1}$ and $x_s e^{-g_3}$ should also be smaller than unity, i.e., $x_s$ can never exceed $c^*{}_1 = e^{g_1}$ or $c^*{}_3 = e^{g_3}$. Thus, for each type of capsid, there is a critical free protein concentration $c_T{}^*$ below which the concentration of capsids is almost zero as the number of subunits in the capsids, $q_T$, is large compared to unity. For CCMV, the basic protein subunits are dimers, so $q_1 = 30$ for the $T = 1$ and $q_3 = 90$ for the $T = 3$ capsid.

Using the equilibrium theory described above, we can now set up the kinetic theory of capsid assembly and disassembly within the framework of CNT.[8] The Gibbs free energy of the formation of an incomplete spherical capsid of the $T$ species containing $n_T = 1, \ldots, q_T$ molecules with a circular rim can be written as

$$\Delta G_T(n_T) = n_T \, \Delta\mu_T + a_T \sqrt{q_T(q_T - n_T)}, \qquad \text{Eq. (S3)}$$

where $a_T = 4\pi R_T \sigma_T / q_T$ is a dimensionless magnitude of the rim energy, with $R_T$ the radius of the shell and $\sigma_T$ the free energy cost per unit length of the rim.[10] $\sigma_T$ can be estimated as $\sigma_T = \frac{-s g_T}{r_T}$ where $s \in [0,1]$ is a geometric factor indicating the average fraction of bonds that a subunit on the rim is missing, which depends on the local coordination number and roughness of the rim. $r_T$ is the effective diameter of a protein unit that is approximated as a disk. Assuming that the surface of a fully formed capsid is covered entirely by capsid proteins, the effective diameter can be written as $r_T = \frac{\sqrt{q_T}}{4 R_T}$.[8] A previous and more detailed study on the line tension of shells composed of Lennard-Jones disks, packed on the surface of sphere, shows that $s \approx 0.3 - 0.4$, with the latter value an upper limit as proteins are more flexible than Jennard-Jones particles - we set $s = 0.3$ in our simulations.[19] Finally, the first term in Eq. (S3) represents the thermodynamic driving force for the assembly or disassembly of capsids. To obtain the (dimensionless) barrier height, $\Delta G_T^*$ for the two $T$ numbers, we calculate the critical nucleus size $n_T^*$, that is, find the value of $n_T$ for which Eq. (S3) is maximal and insert this into Eq. (S3) to obtain

$$\Delta G_{as,T}^* = \Delta G_T^0 \left( \sqrt{\Gamma_T^2 + 1} - \Gamma_T \right), \qquad \text{Eq. (S4)}$$

where $\Delta G_T^0 \equiv q_T a_T / 2$ is the maximum barrier height and $\Gamma_T \equiv -\Delta\mu_T / a_T$ is a measure for the degree of super- or undersaturation for the species $T$. We note that $\Gamma_T$ depends on the concentration of a capsid.

In the process of assembly of a capsid, subunits can attach to and detach from the growing shell through a sequence of what we presume to be reversible kinetic steps. Within the classical nucleation theory, the rate of capsid assembly is limited by the rate of the formation of the energetically most unfavorable critical nucleus through the Boltzmann weight $\Delta G_T^*$ that acts as a kinetic bottleneck.

The steady-state nucleation rate for *association* reads[10]

$$J_{as,T} = x_s \, \nu^*_T Z_T \, exp\left(-\Delta G^*_{as,T}\right), \qquad \text{Eq. (S5)}$$

where $Z_T = \sqrt{-\frac{1}{2\pi}(\partial^2 \Delta G_T/\partial n^2)_{n=n_T^*}} = \sqrt{\frac{a_T}{q_T \pi}}(1+\Gamma_T^2)^{3/4}$ is the so-called Zeldovich factor that describes the sharpness of the free energy barrier and that may be interpreted as a measure of the lifetime of the critical nucleus of size $n_T^*$.[13] The attempt or attachment frequency $\nu^*_T$ of the monomers attaching to the critical nucleus depends on the mode of attachment, and may be , *e.g.*, a function of the diffusivity and concentration of the free monomers, the size of critical nucleus, and on some internal molecular time scale associated with the docking process that may depend on conformational switching.[20] For simplicity, we assume that it does not depend on the size of the clusters nor on the concentration.

To model the disassembly process, we presume that the initial state constitutes a fully formed capsid. Within classical nucleation theory, the rate of disassembly of a capsid is limited by dissociation of the critical nucleus, *i.e.*, it depends on the time required to jump over the height of the free energy barrier $\Delta G_{dis,T}^*$ albeit in the opposite direction from the assembly process. The nucleation rate for disassembly of complete capsids of species $T$ reads

$J_{dis,T} = x_T \, \nu^*_T Z_T \, exp \, (-\Delta G^*_{dis,T})$ ,  Eq. (S6)

where $\Delta G_{dis,T}^*$ represents the free energy barrier for the disassembly of a shell to form monomers. Notice that the dissociation rate depends on $x_T$, the capsid concentrations of species $T = 1,3$. We shall presume that the attachment frequency associated with the association process is the same as that of the dissociation process, as it describes the same process and we presume it to be independent of the size of the critical nucleus.[21]

Because capsids with different $T$ numbers have different radius of curvature, we do not allow for a direct transition from one $T$ number to another one. In our allowed reaction path pathway, growth or disassembly can only proceed by the shedding or docking of individual protein subunits, which for CCMV constitute coat protein dimers. This is not a far-fetched reaction path, as our experiments show no indication of partially disassembled $T = 3$ particles spontaneously morphing into $T = 1$ particles, or vice versa, $T = 1$ particles opening up to absorb subunits and growing into a $T = 3$ particles. Hence, we presume that, first, one type of capsid disassembles into dimers, second, free dimers reassemble into different capsid sizes following their corresponding assembly nucleation rates.

Presuming that kinetic processes are sufficiently slow to allow us to use the expressions for steady-state nucleation rates for association and dissociation, *i.e.*, presume a quasi steady state to hold, the set of equations describing the concentration of dimers, $T = 1$ and $T = 3$ capsids in our system can finally be expressed as follows,

$$\frac{dx_s}{dt} = -q_1 J_{as,1} - q_3 J_{as,3} + q_1 J_{dis,1} + q_3 J_{dis,3} ,\qquad\text{Eq. (S7)}$$

and

$$\frac{dx_T}{dt} = J_{as,T} - J_{dis,T}.\qquad\text{Eq. (S8)}$$

The quantities on the left-hand sides of Eqs. (S7 and S8) represent time derivatives of the concentrations of the three species in our model. The terms on the right-hand sides are due to the changes in the formation or dissociation of capsids, where we let all concentrations be time dependent yet obey mass conservation at all times. We solve the above system of equations numerically, using an explicit forward Euler method with adaptive time steps. (See SI 3.2 for more information.)

From the solutions of these equations, we calculate the fraction of dimers in each type of capsid compared to the total number of dimers *in all capsids*,

$$f_T = \frac{q_T x_T}{\sum_{i=1,3} q_i x_i}\qquad\text{Eq. (S9)}$$

where $T = 1,3$ depending on the $T$ number. In general, the time steps in the simulations depend on various parameters, such as the attempt frequency, binding energies and size of the capsids. In order to fit the theory to the experimental data, we choose one experimental data point in the time series of the disassembly of $T = 3$ and the assembly of $T = 1$, and use it as our reference point. When we find the same ratio of proteins in the two kinds of capsid in our simulations as that in the experiments, we set the time in the simulations equal to the time in the experiments. Next, we rescale all other simulation data points accordingly. We repeat this process for the disassembly of $T = 1$ and the assembly of $T = 3$. To check the robustness of this technique, we consider different data points as the reference point. The agreement between theory and experiments does not depend strongly on the reference point that we take. (See the SI 3.4 for more information.)

It is important to realize that the time that it takes to change the $pH$ and the salt concentration of the buffer solution might not be exactly the same in each experiment. In addition, the lag time for assembly and disassembly of capsids with different sizes are arguably different. Therefore it is difficult to pinpoint the actual "time zero" for each individual experiment. In order to deal with this uncertainty, we start collecting data 30 minutes after the experiment commences. We also assume that the lag times are negligible on the time scale of the experiment, thus we ignore the first phase of assembly in CNT in which capsids have not started to assemble or disassemble.[10]

## 3.2 Numerics

The kinetics equations predicted by CNT (Eqs. S7 and S8) are solved by using finite difference methods. Assembly and disassembly nucleation rates at the beginning of the simulation are determined by the initial conditions. The concentrations of capsids and free dimers are calculated at each time step, using the values and nucleation rates at the previous time step. Hence, our time-stepping equations read:

$$x_s^{t+\Delta t} = x_s^t + (-q_1 J^t_{as,1} - q_3 J^t_{as,3} + q_1 J^t_{dis,1} + q_3 J^t_{dis,3}) \times \Delta t ,\qquad \text{Eq. (S10)}$$

and

$$x_T^{t+\Delta t} = x_T^t + (J^t_{as,T} - J^t_{dis,T}) \times \Delta t,\qquad \text{Eq. (S11)}$$

where $x_s^t$ and $x_T^t$ are the concentrations of free dimers and capsids of size $T$ at time $t$; $q_1$ and $q_3$ represent the number of subunits in each fully formed capsid of size $T = 1$ and $T = 3$. As already alluded to, we assume that dimers are our building blocks, so that $q_1 = 30$ and $q_3 = 90$. The instantaneous steady-state assembly and disassembly nucleation rates of capsid size T at time $t$ are $J^t_{as,T}$ and $J^t_{dis,T}$, respectively.

The nucleation rates depend highly on the free dimer concentration. In order to speed up the simulation, we use an adaptive time step such that a maximum of $0.01\%$ of free dimers at time $t$ can be consumed by growing shells. Similarly, a maximum of $0.01\%$ of free dimers can be released by dissociated shells at time $t$. In other words,

$$\Delta t = \frac{0.0001\, x_s}{Max(J^t_1, J^t_3)}\qquad \text{Eq. (S12)}$$

where $J^t_T = |J^t_{as,T} - J^t_{dis,T}|$ is the absolute assembly or disassembly rate of capsid size $T$. The simulation continues until full depletion of the unfavorable capsid size.

From equilibrium theory and experimental observations, we have to assume that there are some free dimers remaining in the solution before the quench, that is, before the induced shift in $pH$ and in salt concentration that on the time scale of the experiment is (virtually) instantaneous. Therefore, we invoke a non-zero value as our initial free dimer concentration. Quickly after starting the simulation, the dimer concentration converges to a fixed concentration relatively close to what must be the smaller critical concentration. Having initially more dimers in the system leads to the fast formation of capsids. On the other hand, a low dimer concentration at the start of the simulation increases the initial disassembly rate. In order to avoid both of these conditions, we choose the initial dimer concentration near the concentration it converges to. It also helps us to avoid any divergence in the simulation as the dissociation rates increase significantly at low dimer concentrations.

Based on the dimer concentration at the end of the experiment, which is relatively close to the critical concentration of the more stable species, we approximate the total dimer concentration is around 10 times larger than the critical concentration. Therefore, the overall protein in the unfavorable capsid we set at $10\ c^*$, where $c^* = Min(c^*_1, c^*_3)$. (See table S4 and S5 for parameter values.)

Due to the universality of the phenomena, the model is capable of reproducing the experimental results by using different binding energies. We decide to fix the binding energy of T=1 in our framework and generate experimental results only by changing the binding energy of T=3. This allows us to have a better comparison between the two types of experiment.

### 3.3 The fraction of protein dimers in free solution and in capsids

In the main text we show results of the fraction of proteins in the two types of capsid, $f_T$, as we find this the most informative. Alternatively, we can also calculate the fraction of dimers in one type of capsid relative to the *total* number of dimers in solution,

$$h_T = \frac{q_T x_T}{c_S},$$ Eq. (S13)

where $T = 1,3$ is as before the T number of the capsid and $c_S$ the overall dimer concentration in solution. Our simulations show that for our choice of parameters the concentration of dimers remains constant during the process (see Figure S5), implying that the increase in the protein fraction in capsids is not due to the assembly of the free dimers initially present in the solution. This agrees with what is seen in the experiments, see Figure S7. This supports our suggestion that one capsid size disassembles into free dimers and that these proteins re-assemble into the other capsid size, and that the fraction $f_T$ is the relevant quantity describing the assembly and disassembly kinetics for the problem in hand.

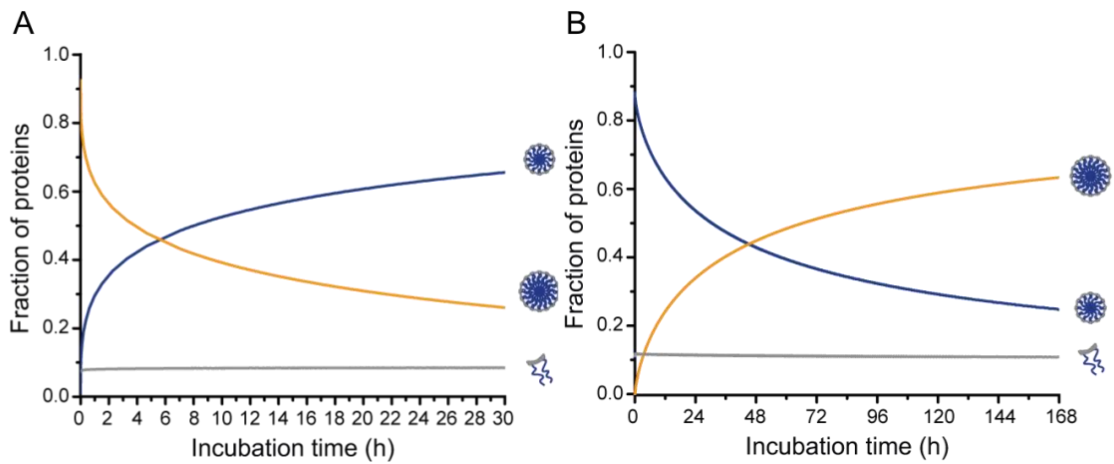

**Figure S5**. The fraction of protein in T=1 (blue line) and T=3 (yellow line) and dimer (gray line) over the total amount of protein in the solution ($h_T$) for the experiments showing A) disassembly of T=1 (represented in the figure 3) and B) disassembly of T=3 (represented in the figure 2). Refer to table S4 and S5 for the parameter values.

### *3.4 Fitting of the simulation results with respect to different reference points*

We have calibrated the simulation results using a reference point in the data series, as mentioned in the caption for figure 2C and 3C. Here we re-calibrated the same simulation data set with respect to a number of reference points to verify the robustness of our fitting procedure. We find that the curve fits depend only relatively weakly on the choice of reference point (Figure S6). Unfortunately, our numerical implementation of Classical Nucleation Theory does not allow us to find the fundamental time scale, that is, the time scale associated with the attempt frequency. In spite of this, we are able to show that the disassembly and assembly of the two different capsid sizes can be explained by CNT.

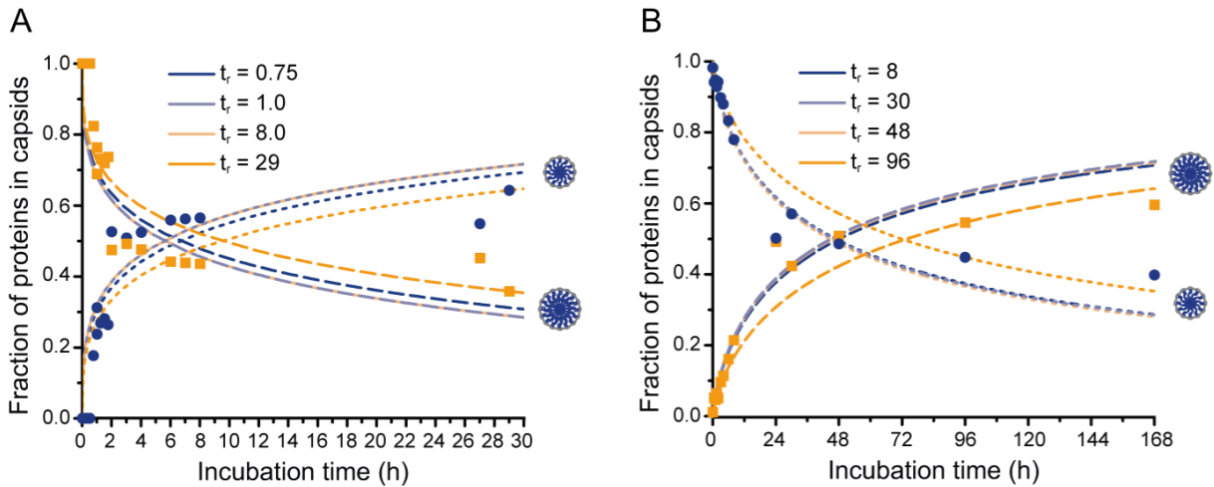

**Figure S6**. The fraction of protein in capsids ($f_T$ as described in Eq S8) of T=1 (short dashed lines) and T=3 (long dashed lines) for A) disassembly of T=3 (represented in the figure 2) and B) disassembly of T=1 (represented in the figure 3) experiments, using different reference times. See table S4 and S5 for parameter values.

## 3.5 Table of parameters

All parameters related to the simulation of disassembly of T=3 and assembly of T=1, and vice versa, discussed in the main text and used in our comparison with the experiments are tabulated below.

**Table S4 – Parameters used in simulation of disassembly of T=3 and assembly of T=1**

| Parameter | Value ( unit) | Description |
|---|---|---|
| $g_1$ | -15 ($k_BT$) | Binding energy of T=1 capsids |
| $g_3$ | -14.7 ($k_BT$) | Binding energy of T=3 capsids |
| $q_1$ | 30 | Number of dimers in fully formed T=1 |
| $q_3$ | 90 | Number of dimers in fully formed T=3 |
| $x_s(t=0)$ | $0.8\, c^*_1$ | Initial dimer concentration |
| $x_1(t=0)$ | 0 | Initial T=1 concentration |
| $x_3(t=0)$ | $\frac{1}{9} c^*_1$ | Initial T=3 concentration |
| $\nu^*$ | 1 ( a.u.) | Critical attempt frequency |
| s | 0.3 | fraction of bonds that a rim protein has fewer than a core protein |
| $t_r$ | 4 ( hours) | Reference time |

**Table S5 – Parameters used in simulation of disassembly of T=1 and assembly of T=3**

| Parameter | Value ( unit) | Description |
| --- | --- | --- |
| $g_1$ | -15 ( $k_BT$) | Binding energy of T=1 capsids |
| $g_3$ | -15.4 ( $k_BT$) | Binding energy of T=3 capsids |
| $q_1$ | 30 | Number of dimers in fully formed T=1 |
| $q_3$ | 90 | Number of dimers in fully formed T=3 |
| $x_s(t=0)$ | $1.3\, c^*_3$ | Initial dimer concentration |
| $x_1(t=0)$ | $\frac{1}{3} c^*_3$ | Initial T=1 concentration |
| $x_3(t=0)$ | 0 | Initial T=3 concentration |
| $\nu^*$ | 1 ( a.u.) | Critical attempt frequency |
| s | 0.3 | fraction of bonds that a rim protein has fewer than a core protein |
| $t_r$ | 48 ( hours) | Reference time |

# 4 Supplemental figures

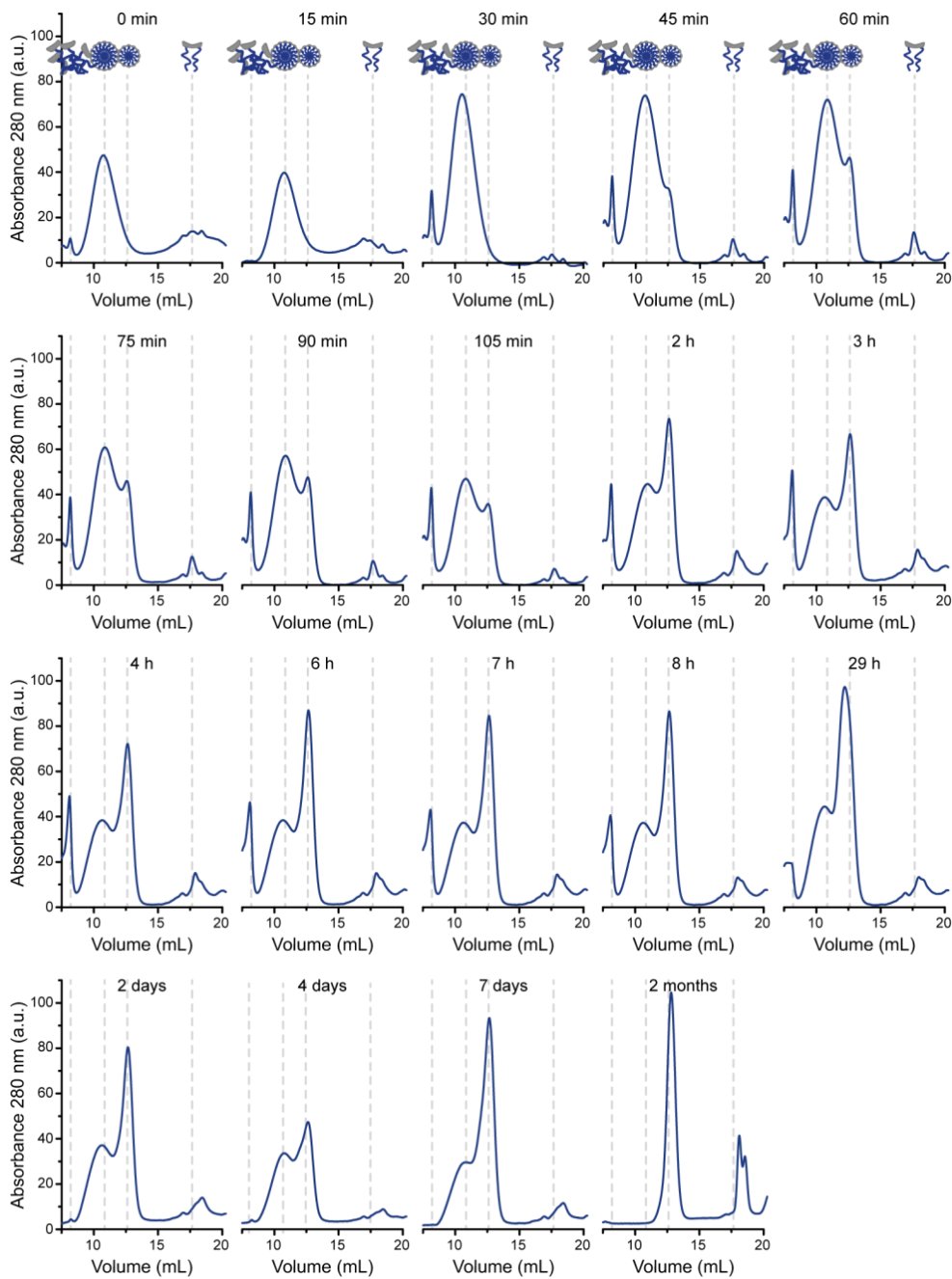

**Figure S7.** SEC analysis of the shift from $T = 3$ to $T = 1$ capsids (dataset 1). SEC chromatograms were measured after the indicated dialysis times to pH 7.5 buffer with 100 mM NaCl at 4 °C. Samples were stabilized by incubation with 0.2 eq. of $Ni^{2+}$ for 50 minutes at room temperature prior to analysis. The results from three separate experiments are combined: 0 to 105 minutes, 2 h to 29 h and 2 days to 2 months.

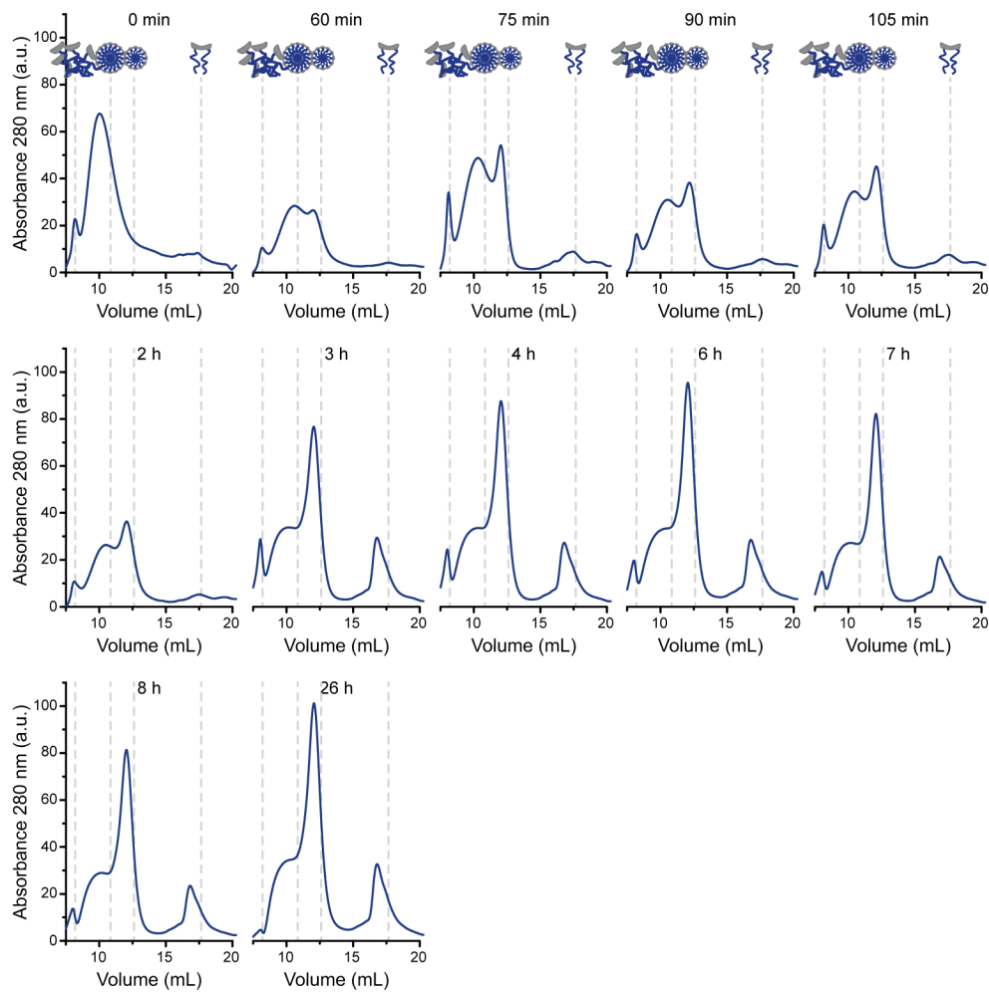

**Figure S8.** SEC analysis of the shift from $T = 3$ to $T = 1$ capsids (dataset 2). SEC chromatograms were measured after the indicated dialysis times to pH 7.5 buffer with 100 mM NaCl at 4 °C. Samples were stabilized by incubation with 0.2 eq. of $Ni^{2+}$ for 50 minutes at room temperature prior to analysis. The results from two separate experiments are combined: 0 to 2 h and 3 h to 26 h.

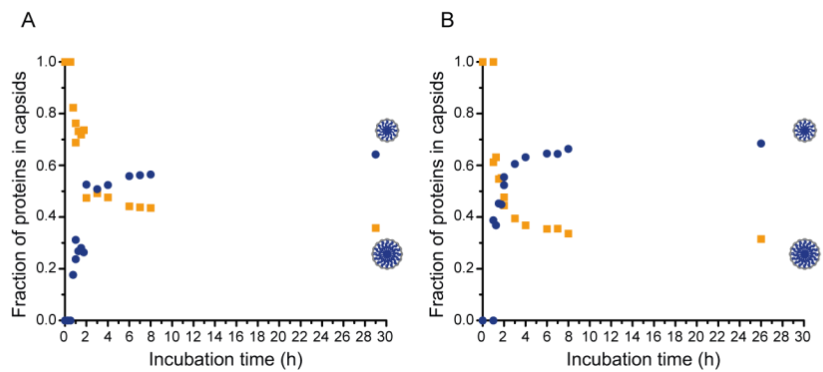

**Figure S9.** Comparison of time curves based on both datasets for the shift from $T = 3$ to $T = 1$ capsids. Protein fractions as $T = 1$ (blue circles) and $T = 3$ (yellow squares) capsids as determined by integration of the SEC chromatograms of Figure S7 (A) and Figure S8 (B).

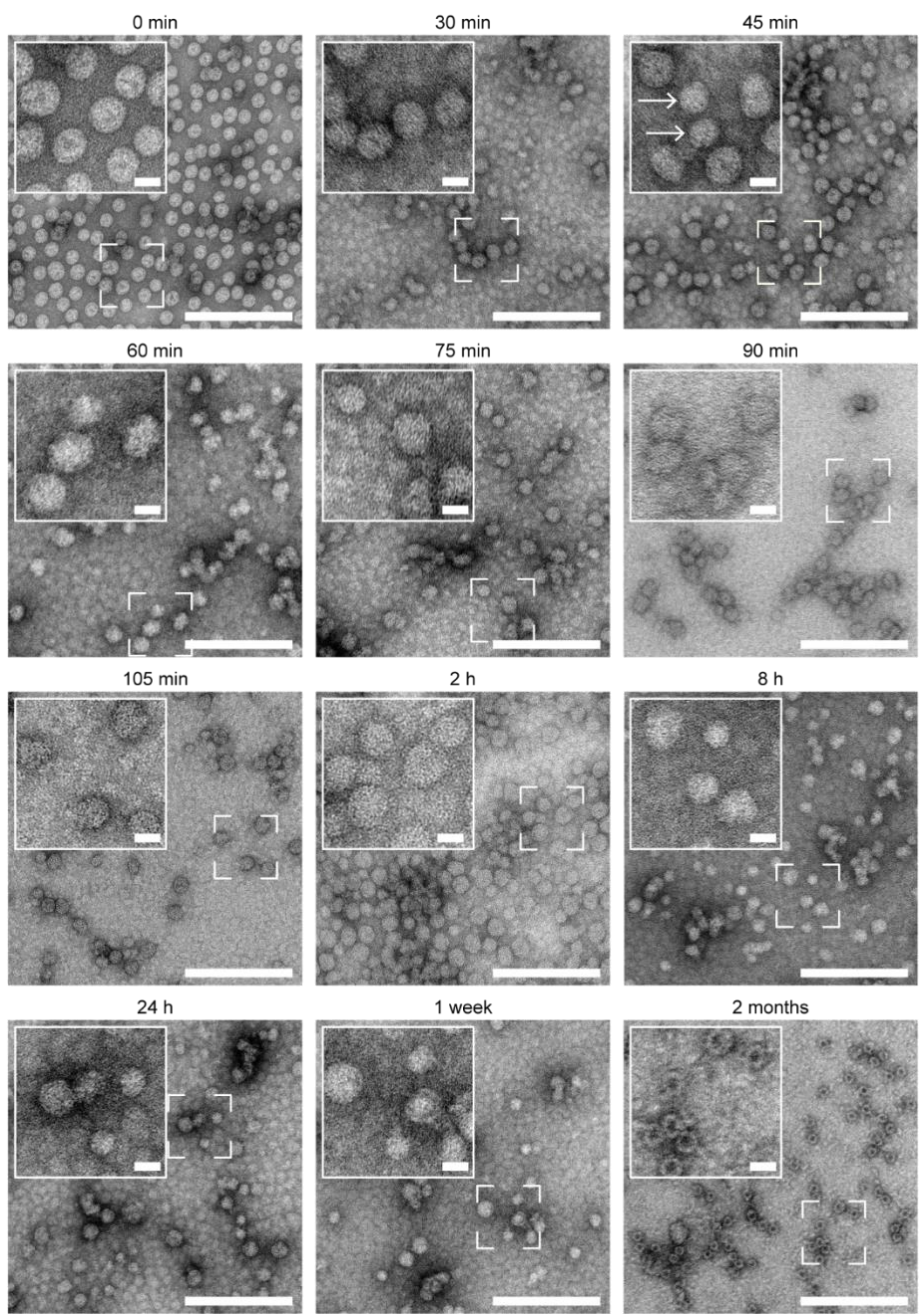

**Figure S10.** TEM analysis of VW1-VW8 ELP-CCMV capsids during size decrease in pH 7.5 buffer with 100 mM NaCl. Samples were taken after the indicated dialysis times at 4 °C and negatively stained with uranyl acetate. Areas indicated with brackets are displayed as zoomed images with $T = 1$ capsids indicated by arrows. Scale bars correspond to 200 nm (main images) and 20 nm (zoomed images).

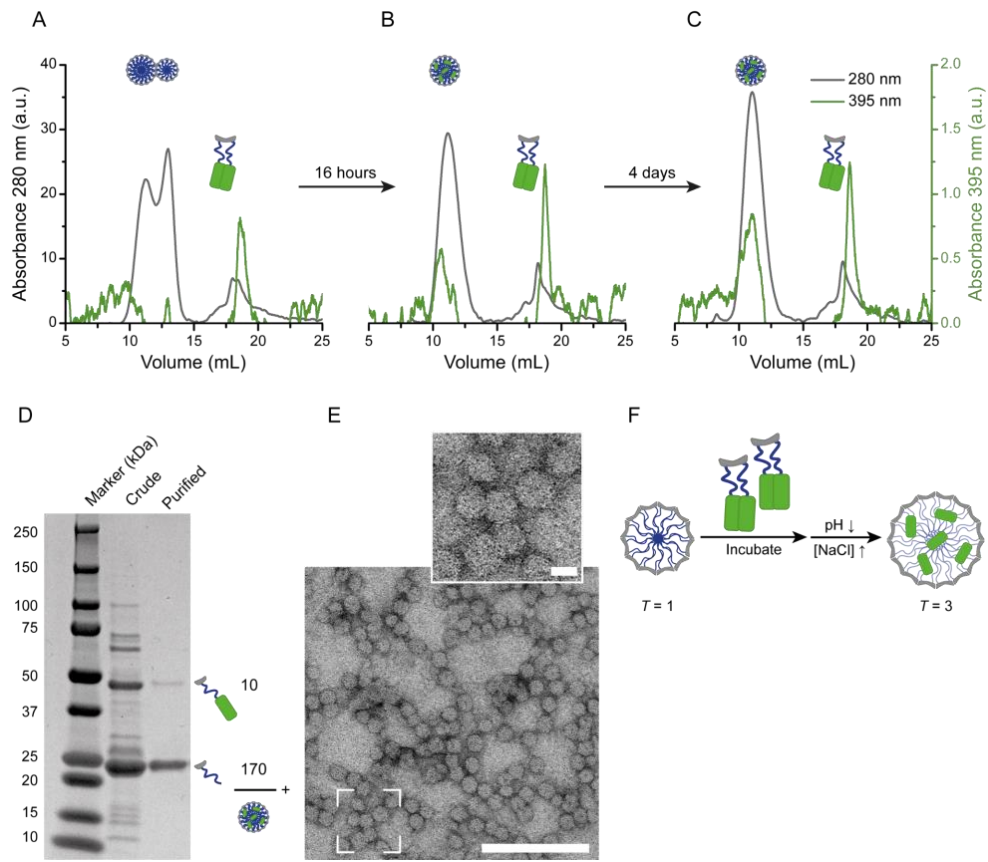

**Figure S11.** Analysis of mEGFP-labeling of *T*=3 VW1-VW8 ELP-CCMV capsids by mixing empty *T*=1 capsids with labeled mEGFP-VW1-VW8 ELP-CCMV dimers. A-C) SEC chromatograms of 25 μM mEGFP-VW1-VW8 ELP-CCMV and 100 μM VW1-VW8 ELP-CCMV capsid mixtures in pH 5.0 buffer with protein absorbance measured at 280 nm (gray curve and axis) and mEGFP absorbance measured at 395 nm (green curve and axis). Mixtures were allowed to mix at 4 °C, pH 5.0 for 90 minutes (A), overnight (B) or 4 days (C). The increase in mEGFP signal indicates that the mEGFP-labeled dimers exchange with non-labeled dimers over time. D) SDS-PAGE analysis of SEC purification after 4 days of exchange. The gel was stained with Coomassie Brilliant Blue. ImageJ quantification is depicted next to the gel. E) Uranyl acetate-stained TEM micrograph of SEC-purified capsids. The area indicated with brackets is displayed as a zoomed image. Scale bars correspond to 200 nm (main image) and 20 nm (zoomed image). F) Schematic of the process we studied: empty *T*=1 capsids are mixed with mEGFP-VW1-VW8 ELP-CCMV dimers, which upon a decrease in pH and increase in NaCl concentration exchange with non-labelled dimers to form labeled *T*=3 capsids.

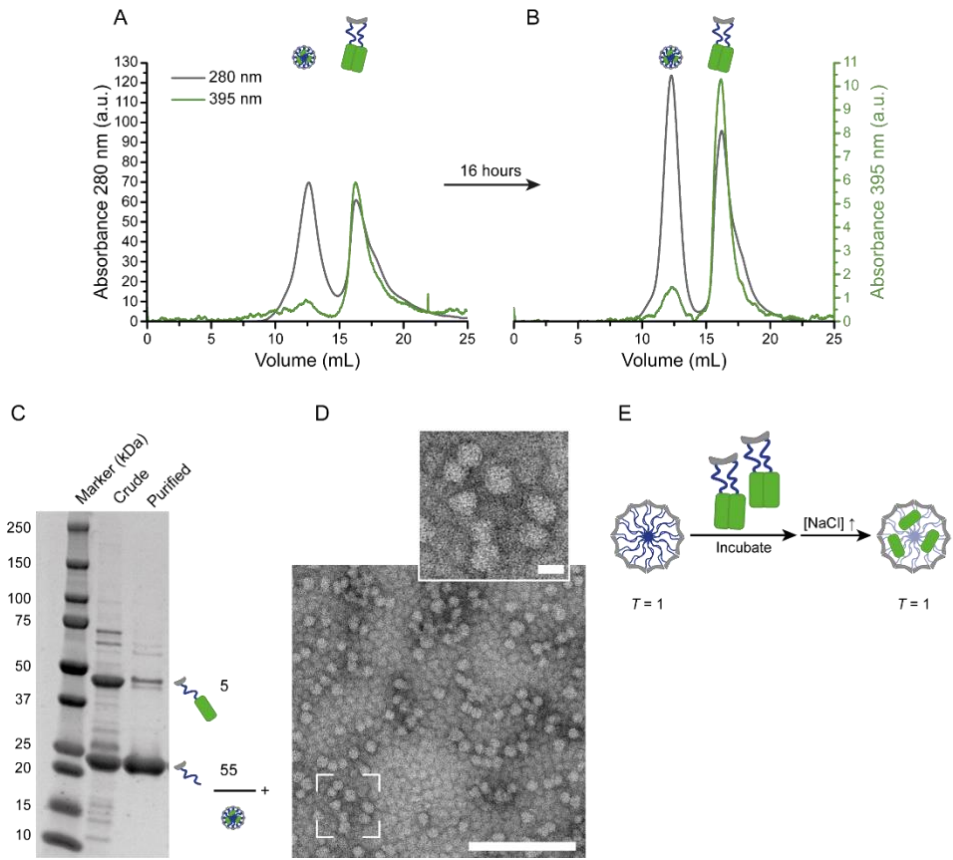

**Figure S12.** Analysis of mEGFP-labeling of *T*=1 VW1-VW8 ELP-CCMV capsids by mixing empty *T*=1 capsids with labeled mEGFP-VW1-VW8 ELP-CCMV dimers. A-B) SEC chromatograms of 25 µM mEGFP-VW1-VW8 ELP-CCMV and 100 µM VW1-VW8 ELP-CCMV capsid mixtures in pH 7.5. 500 mM NaCl buffer with protein absorbance measured at 280 nm (gray curve and axis) and mEGFP absorbance measured at 395 nm (green curve and axis). Mixtures were allowed to mix at 4 °C, pH 7.5, 500 mM NaCl for 90 minutes (A) or overnight (B). The increase in mEGFP signal indicates that the mEGFP-labeled dimers exchange with non-labeled dimers over time. C) SDS-PAGE analysis of SEC purification after overnight exchange. The gel was stained with Coomassie Brilliant Blue. ImageJ quantification is depicted next to the gel. D) Uranyl acetate-stained TEM micrograph of SEC-purified capsids. The area indicated with brackets is displayed as a zoomed image. Scale bars correspond to 200 nm (main image) and 20 nm (zoomed image). E) Schematic of the process we studied: empty *T*=1 capsids are mixed with mEGFP-VW1-VW8 ELP-CCMV dimers, which upon an increase in NaCl concentration exchange with non-labelled dimers to form labeled *T*=1 capsids.

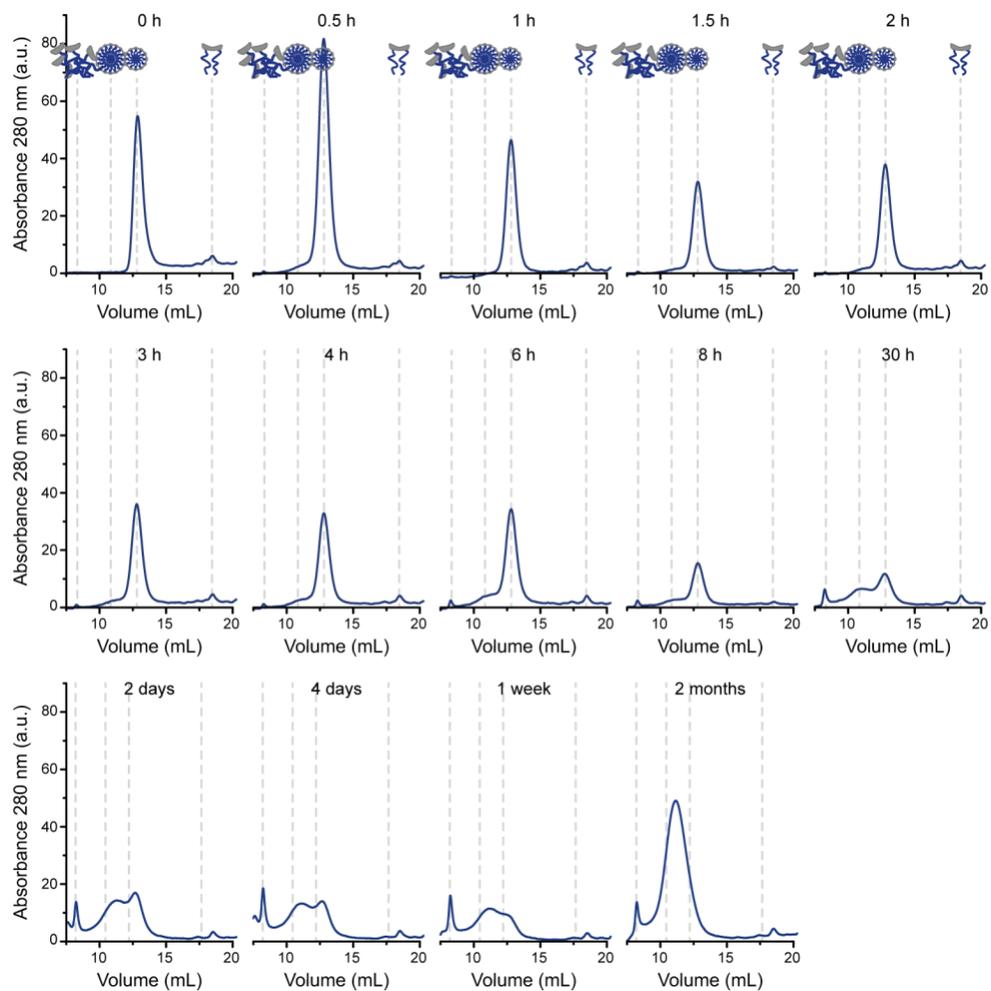

**Figure S13.** SEC analysis of the shift from $T = 1$ to $T = 3$ capsids (dataset 1). SEC chromatograms were measured after indicated dialysis times to pH 5.0 buffer with 500 mM NaCl at 4 °C. The results from two separate experiments are combined: 0 to 30 h and 2 days to 2 months.

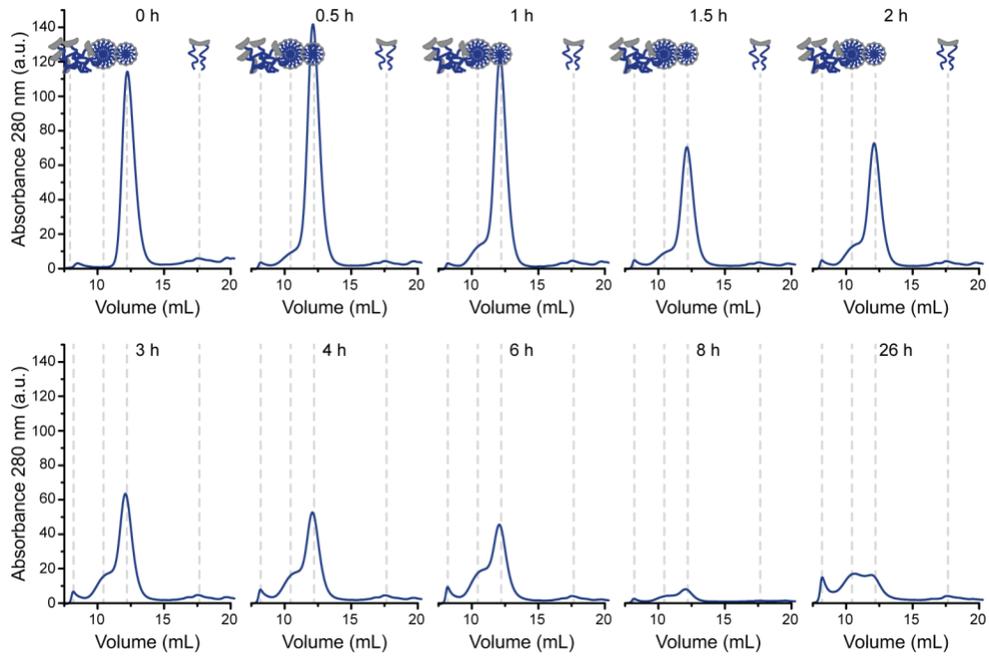

**Figure S14.** SEC analysis of the shift from $T = 1$ to $T = 3$ capsids (dataset 2). SEC chromatograms were measured after indicated dialysis times to pH 5.0 buffer with 500 mM NaCl at 4 °C.

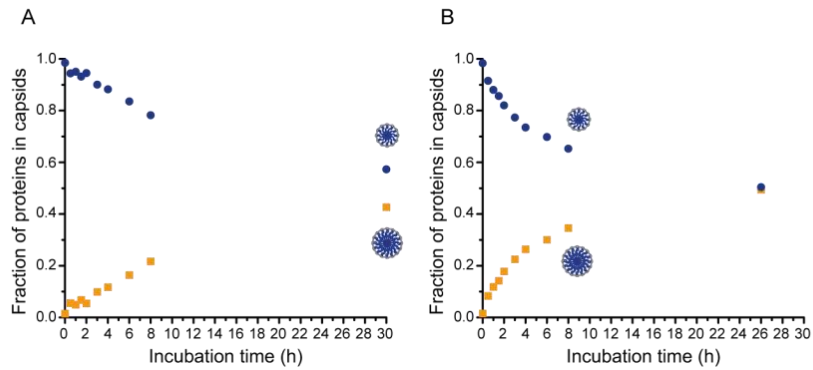

**Figure S15.** Comparison of time curves based on both datasets for the shift from $T = 1$ to $T = 3$ capsids. Protein fractions as $T = 1$ (blue circles) and $T = 3$ (yellow squares) capsids as determined by integration of the SEC chromatograms of Figure S11 (A) and Figure S12 (B).

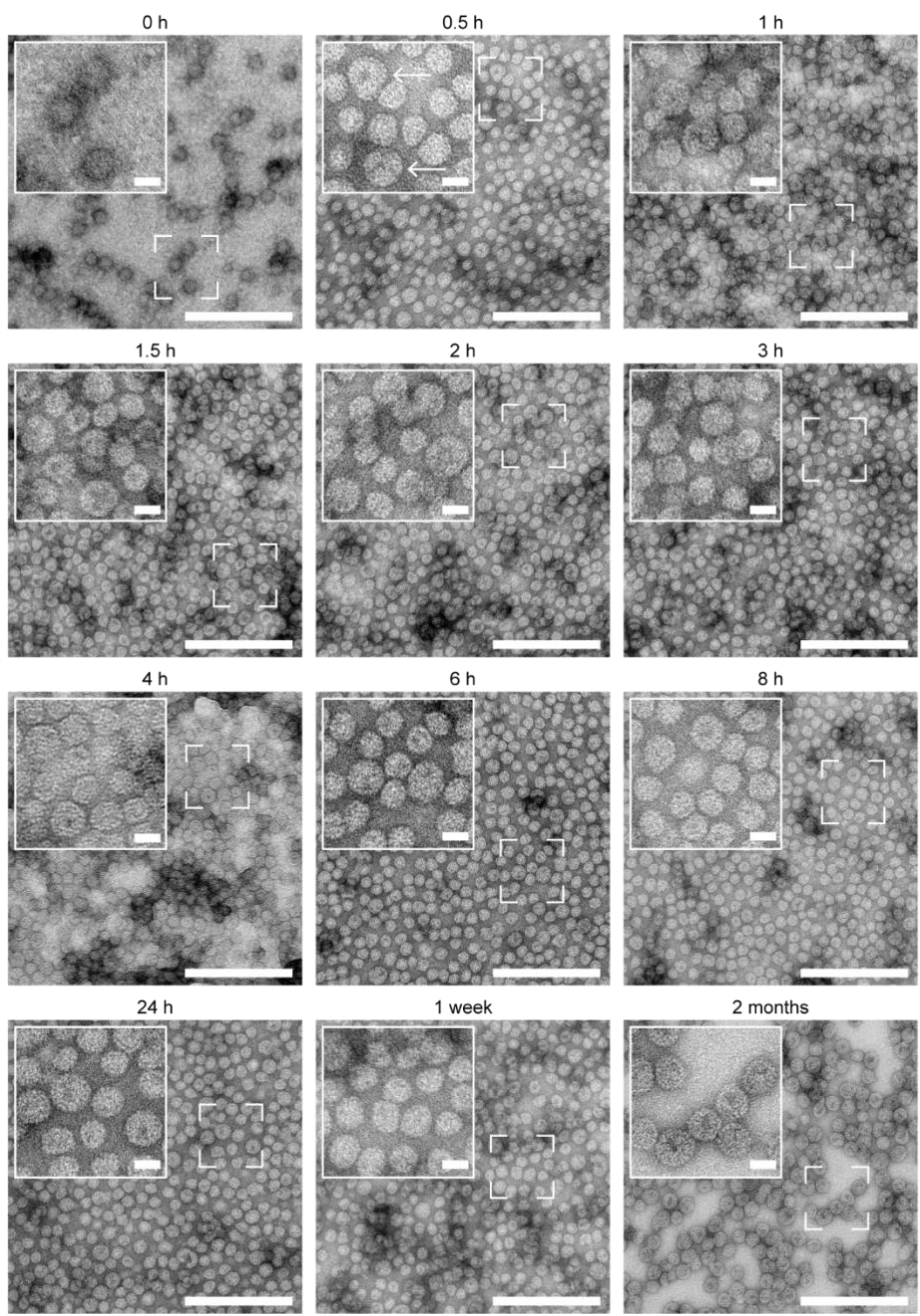

**Figure S16.** TEM analysis of VW1-VW8 ELP-CCMV capsids during size increase in pH 5.0 buffer. Samples were taken after the indicated dialysis times at 4 °C and negatively stained with uranyl acetate. Areas indicated with brackets are displayed as zoomed images with $T$ = 3 capsids indicated by arrows. Scale bars correspond to 200 nm (main images) and 20 nm (zoomed images).

# 5 References

(1) Gill, S. C.; von Hippel, P. H. Calculation of Protein Extinction Coefficients from Amino Acid Sequence Data. *Anal. Biochem.* **1989**, *182* (2), 319–326.

https://doi.org/10.1016/0003-2697(89)90602-7.
(2) Schoonen, L.; Maas, R. J. M.; Nolte, R. J. M.; van Hest, J. C. M. Expansion of the Assembly of Cowpea Chlorotic Mottle Virus towards Non-Native and Physiological Conditions. *Honor Profr. Ben Feringa 2016 Tetrahedron Prize Creat. Org. Chem. Dyn. Funct. Mol. Syst.* **2017**, *73* (33), 4968–4971. https://doi.org/10.1016/j.tet.2017.04.038.
(3) van Eldijk, M. B.; Wang, J. C.-Y.; Minten, I. J.; Li, C.; Zlotnick, A.; Nolte, R. J. M.; Cornelissen, J. J. L. M.; van Hest, J. C. M. Designing Two Self-Assembly Mechanisms into One Viral Capsid Protein. *J. Am. Chem. Soc.* **2012**, *134* (45), 18506–18509. https://doi.org/10.1021/ja308132z.
(4) Schoonen, L.; Pille, J.; Borrmann, A.; Nolte, R. J. M.; van Hest, J. C. M. Sortase A-Mediated N-Terminal Modification of Cowpea Chlorotic Mottle Virus for Highly Efficient Cargo Loading. *Bioconjug. Chem.* **2015**, *26* (12), 2429–2434. https://doi.org/10.1021/acs.bioconjchem.5b00485.
(5) Timmermans, S. B. P. E. Artificial Organelles Based on Hybrid Protein Nanoparticles. PhD Thesis, Chemical Engineering and Chemistry, 2022.
(6) Minten, I. J.; Wilke, K. D. M.; Hendriks, L. J. A.; van Hest, J. C. M.; Nolte, R. J. M.; Cornelissen, J. J. L. M. Metal-Ion-Induced Formation and Stabilization of Protein Cages Based on the Cowpea Chlorotic Mottle Virus. *Small Weinh. Bergstr. Ger.* **2011**, *7* (7), 911–919. https://doi.org/10.1002/smll.201001777.
(7) Bruinsma, R. F.; Wuite, G. J. L.; Roos, W. H. Physics of Viral Dynamics. *Nat. Rev. Phys.* **2021**, 1–16. https://doi.org/10.1038/s42254-020-00267-1.
(8) Michaels, T. C. T.; Bellaiche, M. M. J.; Hagan, M. F.; Knowles, T. P. J. Kinetic Constraints on Self-Assembly into Closed Supramolecular Structures. *Sci. Rep.* **2017**, *7* (1), 12295. https://doi.org/10.1038/s41598-017-12528-8.
(9) Zandi, R.; Dragnea, B.; Travesset, A.; Podgornik, R. On Virus Growth and Form. *Phys. Rep.* **2020**, *847* (C). https://doi.org/10.1016/j.physrep.2019.12.005.
(10) Zandi, R.; Schoot, P. van der; Reguera, D.; Kegel, W.; Reiss, H. Classical Nucleation Theory of Virus Capsids. *Biophys. J.* **2006**, *90* (6), 1939–1948. https://doi.org/10.1529/biophysj.105.072975.
(11) Harms, Z. D.; Selzer, L.; Zlotnick, A.; Jacobson, S. C. Monitoring Assembly of Virus Capsids with Nanofluidic Devices. *ACS Nano* **2015**, *9* (9), 9087–9096. https://doi.org/10.1021/acsnano.5b03231.
(12) Moerman, P.; van der Schoot, P.; Kegel, W. Kinetics versus Thermodynamics in Virus Capsid Polymorphism. *J. Phys. Chem. B* **2016**, *120* (26), 6003–6009. https://doi.org/10.1021/acs.jpcb.6b01953.
(13) Zlotnick, A. To Build a Virus Capsid: An Equilibrium Model of the Self Assembly of Polyhedral Protein Complexes. *J. Mol. Biol.* **1994**, *241* (1), 59–67. https://doi.org/10.1006/jmbi.1994.1473.
(14) Keef, T.; Micheletti, C.; Twarock, R. Master Equation Approach to the Assembly of Viral Capsids. *J. Theor. Biol.* **2006**, *242* (3), 713–721. https://doi.org/10.1016/j.jtbi.2006.04.023.
(15) Rapaport, D. C. Role of Reversibility in Viral Capsid Growth: A Paradigm for Self-Assembly. *Phys. Rev. Lett.* **2008**, *101* (18), 186101. https://doi.org/10.1103/PhysRevLett.101.186101.
(16) Berthet-Colominas, C.; Cuillel, M.; Koch, M. H. J.; Vachette, P.; Jacrot, B. Kinetic Study of the Self-Assembly of Brome Mosaic Virus Capsid. *Eur. Biophys. J.* **1987**, *15* (3), 159–168. https://doi.org/10.1007/BF00263680.
(17) Asor, R.; Schlicksup, C. J.; Zhao, Z.; Zlotnick, A.; Raviv, U. Rapidly Forming Early Intermediate Structures Dictate the Pathway of Capsid Assembly. *J. Am. Chem. Soc.* **2020**, *142* (17), 7868–7882. https://doi.org/10.1021/jacs.0c01092.
(18) Zandi, R.; van der Schoot, P. Size Regulation of Ss-RNA Viruses. *Biophys. J.* **2009**, *96* (1), 9–20. https://doi.org/10.1529/biophysj.108.137489.

(19) Luque, A.; Reguera, D.; Morozov, A.; Rudnick, J.; Bruinsma, R. Physics of Shell Assembly: Line Tension, Hole Implosion, and Closure Catastrophe. *J. Chem. Phys.* **2012**, *136* (18), 184507. https://doi.org/10.1063/1.4712304.
(20) Kashchiev, D. In *Nucleation*; Elsevier, 2000; pp 113–290.
(21) Luque Santolaria, A. *Structure, Mechanical Properties, and Self-Assembly of Viral Capsids*; Universitat de Barcelona, 2011.